%% file: main.tex
\documentclass[twocolumn]{aastex631}

\usepackage[utf8]{inputenc}
\usepackage{longtable}
\usepackage[flushleft]{threeparttable}
\usepackage{tabularx}
\usepackage{multirow}
\usepackage{graphicx}
\usepackage{amsmath,amssymb, fixmath}
\usepackage{xcolor}
\usepackage{makecell}  
\usepackage{color,colortbl}
\usepackage{epstopdf}
\usepackage{hyperref}
\usepackage{url}
\usepackage{subfigure}
\usepackage{rotating}
\usepackage{enumitem}\setlist[description]{font=\textendash\enskip\scshape\bfseries}
\usepackage[normalem]{ulem}
\usepackage[multiple]{footmisc}
\usepackage{bigints}
\usepackage{soul}
\usepackage{etoolbox}
\usepackage{booktabs}

\usepackage[nolist]{acronym}
\input{acronyms}

\makeatletter
\patchcmd{\@footnotetext}{\footnotesize}{\scriptsize}{}{}
\makeatother

\makeatletter
\newcommand\footnoteref[1]{\protected@xdef\@thefnmark{\ref{#1}}\@footnotemark}
\makeatother

\graphicspath{{figures/}}

\definecolor{Gray}{gray}{0.9}
\definecolor{orange}{rgb}{0.9,0.5,0}
\definecolor{navyblue}{RGB}{0,0,128}
\definecolor{darkbrown}{RGB}{101,67,33}
\definecolor{darkmagenta}{RGB}{139,0,139}

\newcommand{\beq}{\begin{equation}}
\newcommand{\eeq}{\end{equation}}
\newcommand{\bdm}{\begin{displaymath}}
\newcommand{\edm}{\end{displaymath}}
\newcommand{\T}[1]{\tilde{#1}}

\newcommand{\orcid}[1]{\href{https://orcid.org/#1}{\textcolor[HTML]{A6CE39}{\aiOrcid}}}

\begin{document}

\title{NMMA–Astro-COLIBRI: An Automated Light-Curve Supernovae Classification Service in the Multi-Survey Era}

\author[0000-0002-9108-5059]{R. Weizmann Kiendrebeogo}
\affiliation{Universit\'e Paris-Saclay – CEA – Irfu, F-91191 Gif-sur-Yvette, France}
\affiliation{Université Côte d’Azur, Observatoire de la Côte d’Azur, CNRS, Laboratoire Artemis, 06300 Nice, France}
\affiliation{Laboratoire de Physique et de Chimie de l’Environnement, Université Joseph KI-ZERBO, Ouagadougou, Burkina Faso}

\author[0009-0003-0039-0483]{Bernardo Cornejo Avila}
\affiliation{Universit\'e Paris-Saclay – CEA – Irfu, F-91191 Gif-sur-Yvette, France}

\author[0009-0005-6643-1473]{Sofia Bisero}
\affiliation{Universit\'e Paris-Saclay – CEA – Irfu, F-91191 Gif-sur-Yvette, France}

\author{Maxime Cellier}
\affiliation{Universit\'e Paris-Saclay – CEA – Irfu, F-91191 Gif-sur-Yvette, France}
\affiliation{Ecole d’ing\'enieurs a\'eronautique et spatiale (IPSA), Paris}

\author{Antoine Ciric}
\affiliation{Universit\'e Paris-Saclay – CEA – Irfu, F-91191 Gif-sur-Yvette, France}

\author[0000-0001-5180-2845]{Ilja Jaroschewski}
\affiliation{Universit\'e Paris-Saclay – CEA – Irfu, F-91191 Gif-sur-Yvette, France}

\author[0009-0009-2025-8256]{Henrik~Rose}
\affiliation{Universität Potsdam, Institut für Physik und Astronomie, Karl-Liebknecht-Str. 24/25, 14476 Potsdam, Germany}

\author{Alexandre Saint-Paul}
\affiliation{Universit\'e Paris-Saclay – CEA – Irfu, F-91191 Gif-sur-Yvette, France}
\affiliation{Ecole d’ing\'enieurs a\'eronautique et spatiale (IPSA), Paris}

\author[0000-0003-1500-6571]{Fabian~Sch\"ussler} \affiliation{Universit\'e Paris-Saclay – CEA – Irfu, F-91191 Gif-sur-Yvette, France}

\correspondingauthor{The Astro-Transients Team}
\email{astro.transients@gmail.com}
\email{weizmann.kiendrebeogo@oca.eu}

\begin{abstract}

The surge in publicly available photometric alerts from wide-field surveys requires automated tools for real-time transient classification. We present \texttt{NMMA--Astro-COLIBRI}, an on-demand Bayesian classification service that couples the Nuclear-physics and Multi-Messenger Astrophysics (NMMA) inference framework to the \emph{Astro-COLIBRI} real-time multi-messenger platform. After the detection of an optical transient, if photometry is available, it is quality-filtered. The filtered photometry is fitted by nested sampling against a user-selected model from a library of eleven supernova templates; results are delivered to every user within minutes. Applying two or more models on the same optical transient, the service reports the corresponding log Bayes factors as a quantitative ranking of competing subtypes. 

We demonstrate the workflow on SN\,2021ugl (ZTF21abotose), a Type\,IIb supernova initially mistaken for a kilonova candidate by automated real-time pipelines, comparing competing supernova and kilonova models. In an early-time configuration using only the first $\sim\!6$~days of photometry in two bands (ZTF $g$ and $r$), so ten days before spectroscopic confirmation, the empirical Type\,IIb template recovers the correct classification, favored over both the kilonova template and the kilonova-mimicking shock-cooling model. In the full 47-day, three-band baseline, it again achieves the highest evidence over every competing supernova and kilonova template. These results highlight the importance of a comprehensive supernova template library for kilonova discrimination in the multi-survey era.

\end{abstract}

\section{Introduction}
\label{sec:intro}

The multi-messenger detection of GW170817 \citep{AbEA2017b}, its kilonova counterpart AT2017gfo \citep{CoFo2017,SmCh2017,AbEA2017f}, and the associated short gamma-ray burst GRB\,170817A \citep{GoVe2017,SaFe2017,AbEA2017e,Savchenko_2017} demonstrated that optical wide-field surveys are essential tools for multi-messenger astrophysics. Yet this single event also revealed the central observational challenge for gravitational-wave follow-up observations: for every kilonova expected in the coming observing runs of the LIGO--Virgo--KAGRA network, wide-field surveys such as the Zwicky Transient Facility \citep[ZTF;][]{Bellm:19:ZTFScheduler, Graham2018} and Asteroid Terrestrial-impact Last Alert System \citep[ATLAS;][]{Tonry_2011, Tonry_2018} detect thousands of \acp{SN} in overlapping sky regions and time windows. This data rate will grow by orders of magnitude with the Vera C.\ Rubin Observatory, whose Legacy Survey of Space and Time \citep[LSST;][]{Ivezic2019} is forecast to detect several million supernovae over its ten-year survey, issuing up to $\sim$10 million transient alerts per night. With the predicted kilonova detection rate remaining as low as $0.43^{+0.44}_{-0.20}$ events per year during the fifth gravitational-wave observing run even with ZTF \citep{Kiendrebeogo_2023}, supernovae vastly outnumber genuine kilonova candidates in any wide-field alert stream. Discriminating between \ac{SN} subtypes rapidly and automatically, without waiting days or weeks for spectroscopic confirmation, is therefore not merely a problem of \ac{SN} science in isolation: it is a \emph{prerequisite} for any credible kilonova search.  

Several \ac{SN} sub-classes are particularly deceptive. Among transients flagged in real-time search as fast-fading candidates by ZTFReST~\citep{Andreoni_2021} over six years of ZTF operations, Type\,IIb supernovae form the single largest class, accounting for 20 of the 66 cataloged impostors \citep{Barna_2025}. Between 30 and 50 percent of Type\,IIb events display a shock-cooling tail following shock breakout in an extended progenitor \citep{Ayala_2025}, with the decline from the shock-cooling peak lasting about a week on
average \citep{Crawford_2025}. In this period the light curve fades rapidly and the spectrum remains largely featureless over the first few days \citep[e.g.][]{Wang_2023, Subrayan_2025}, similar to kilonovae \citep{Pian_2017}. \citet{Ackley_2026} estimate that such tails outnumber binary neutron star mergers by a factor of 30 to 300 in untriggered searches; SN\,2025ulz, the only candidate to pass ZTF vetting in the localization region of S250818k \citep{2025GCN.41437....1L}, is the most recent illustration \citep{Franz_2025}. Discriminating these events photometrically, before the $^{56}$Ni rebrightening resolves the ambiguity, requires a model library well beyond the handful of templates used by earlier automated frameworks. \citet{Barna_2024} recovered shock-cooling light curves correctly in at most 36\% of simulated cases with four models, and obtained an odds ratio of only 1.55 between the shock-cooling and kilonova models for ZTF21abotose/SN\,2021ugl \citep{Ridley_2021}. We adopt this Type\,IIb SN as our proof-of-concept case study.

To address this challenge, we present \texttt{NMMA--Astro-COLIBRI}\footnote{\scriptsize \url{https://nmma.live/}\label{docs-link}}, an on-demand Bayesian classification service for the rapid photometric classification of SNe, integrated into the \emph{Astro-COLIBRI} real-time multi-messenger platform \citep{Reichherzer_2021, 2023Galax..11...22R, schussler2025astrocolibricomprehensiveplatformrealtime, schussler2025astrocolibriempoweringcitizenscientists,avila2026enablingrealtimemultimessengerfollowup, avila2026astrocolibriinnovativeplatformrealtime}. The service fits photometric light curves from ZTF, ATLAS, and LSST against a library of eleven \ac{SN} models spanning the Type\,Ia, stripped-envelope, core-collapse, and shock-cooling sub-classes, together with five kilonova models used as an adversarial benchmark (Section~\ref{sec:nmma}). Using the Nuclear-physics and Multi-Messenger Astrophysics (NMMA) Bayesian inference framework \citep{Pang_2023,Rose:2026}, it delivers best-fit light curves, corner plots (showing the one- and two-dimensional marginalized posterior distributions of the sampled parameters; \citealt{corner}), and Bayesian evidences to every \emph{Astro-COLIBRI}\footnote{\scriptsize \url{https://astro-colibri.science/}} client within minutes of submission.

NMMA\footnote{\scriptsize \url{https://nuclear-multimessenger-astronomy.github.io/nmma/}} was previously integrated into the SkyPortal science platform \citep{Coughlin_2023}, enabling user-selectable light-curve classification of kilonovae, supernovae, and GRB afterglows. It was accessible through SkyPortal and its Fritz instance\footnote{\scriptsize\url{https://github.com/fritz-marshal/fritz}} to the time-domain community, building in part on an earlier, kilonova-focused NMMA--SkyPortal pipeline whose results were accessible only to members of the dedicated SkyPortal group \citep{kiendrebeogo:tel-04796327}. Subsequently, \citet{Barna_2024} coupled NMMA with the ZTFReST infrastructure \citep{Andreoni_2021} to deliver automated daily fits of ZTF fast-transient candidates on a high-performance computing cluster. Here we extend this line of work to the \emph{Astro-COLIBRI} platform, open to any registered user without group membership, in two directions:

\begin{enumerate}
  \item multi-survey photometry from the \emph{Astro-COLIBRI} archive, combining ZTF, ATLAS, and LSST bands in a single standardized pipeline.
  \item seamless delivery of classification results directly inside the \emph{Astro-COLIBRI} mobile and web clients, immediately accessible to any registered user browsing the transient's page, without requiring access to a separate data-management platform.
\end{enumerate}

This paper is structured as follows. Section~\ref{sec:platform} reviews the \emph{Astro-COLIBRI} platform, its integration with NMMA, and the full analysis pipeline. Section~\ref{sec:filtering} describes how transients are selected and submitted for analysis. Section~\ref{sec:nmma} describes the NMMA framework and the \ac{SN} and kilonova models implemented here. Section~\ref{sec:results} demonstrates the workflow on SN\,2021ugl. We discuss limitations and future directions in Section~\ref{sec:discussion}, concluding in Section~\ref{sec:conclusion}.

\section{Integrated Data Management in Astro-COLIBRI}
\label{sec:platform}

\begin{figure*}
\centering
\includegraphics[width=\textwidth]{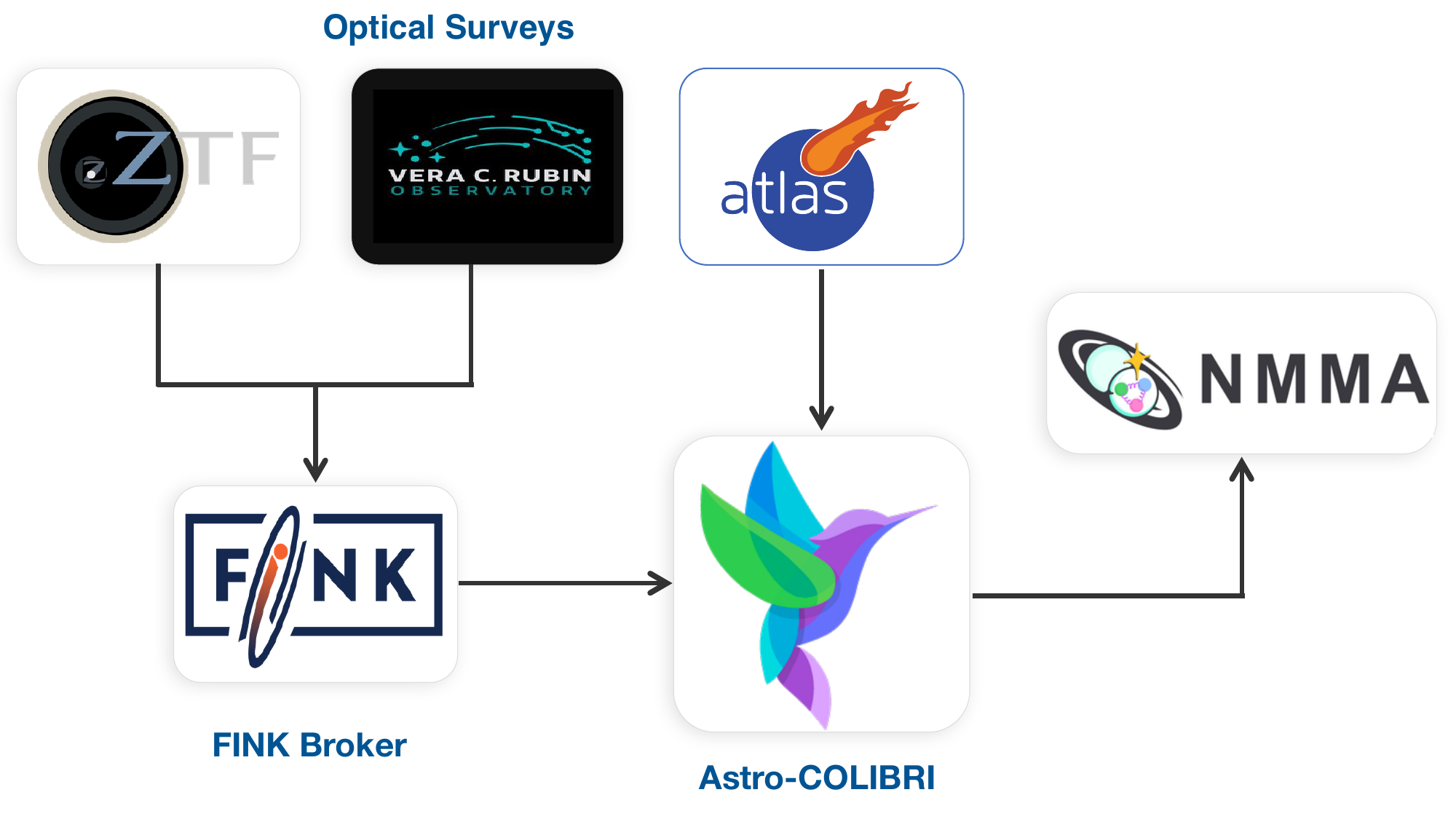}
\caption{Overview of the \texttt{NMMA--Astro-COLIBRI} pipeline. 
        Photometric alerts from ZTF and LSST are ingested via the Fink broker, while ATLAS data flow directly into the \emph{Astro-COLIBRI} archive. A user selects a transient of interest through the \emph{Astro-COLIBRI} interface, triggering automated Bayesian inference with NMMA against a library of eleven supernova and five kilonova models.}
\label{fig:astro-colibri}
\end{figure*}

\textbf{Astro-COLIBRI}---\emph{Astro-COLIBRI} (COincidence LIBrary for Real-time Inquiry) is a real-time multi-messenger platform that ingests, filters, and contextualizes transient alerts from a wide range of astrophysical observatories and brokers \citep{Reichherzer_2021, 2023Galax..11...22R}. Its architecture comprises a RESTful API, static and real-time databases, a cloud-based push-notification system, and web and mobile clients for iOS and Android. Tens of thousands of registered users, including professional and amateur astronomers, receive real-time push notifications and can query the platform through its public API or graphical interface.

A key feature of the platform is its optical light-curve archive, which aggregates photometric data from multiple surveys, including ATLAS, ZTF, LSST, the All-Sky Automated Survey for SuperNovae \citep[ASAS-SN;][]{Shappee_2014}, the American Association of Variable Star Observers \citep[AAVSO;][]{1993Ap&SS.210..137P}, and the Réseau Amateurs Professionnels pour les Alertes Scientifiques\footnote{\scriptsize \url{https://rapas.imcce.fr/}} (RAPAS), into a single standardized repository. Photometry is stored in this archive as per-source CSV files, the format consumed directly by the \texttt{NMMA--Astro-COLIBRI} pipeline; an IVOA-compliant VOTable representation of the same data is also exposed for interoperability with external virtual-observatory tools, but plays no role in the classification pipeline itself (Section~\ref{sec:filtering}). While the full archive spans this broad set of instruments, the current \texttt{NMMA--Astro-COLIBRI} pipeline operates on the three surveys for which NMMA filter responses are implemented: ZTF, ATLAS, and LSST. The cloud infrastructure and maintenance of the NMMA analysis API are provided by \emph{Astro-COLIBRI}, so that fitting results are accessible to all clients without requiring dedicated computing resources on the user side. Figure~\ref{fig:astro-colibri} shows the overall pipeline, from alert ingestion through Bayesian classification to result delivery.

\textbf{Analysis platform}---The \texttt{NMMA--Astro-COLIBRI} service extends the platform by coupling the NMMA Bayesian inference framework \citep{Pang_2023,Rose:2026} to the \emph{Astro-COLIBRI} light-curve archive. Having selected a transient of interest (Section~\ref{sec:filtering}), a user submits a fitting job by choosing one of the eleven \ac{SN} models or five kilonova models described in Section~\ref{sec:nmma}. The photometry CSV is retrieved from the archive, quality-filtered, and converted to the NMMA data format (Section~\ref{sec:filtering}). 

The job is registered as \texttt{pending} in MongoDB and executed asynchronously in a background thread on the server side, leaving the client free to poll the job status at ten-second intervals via the REST API. Once the nested-sampling inference completes, the best-fit parameters, light-curve plots, and corner plots are uploaded to cloud storage and become immediately retrievable by the client. Each user's job history is also persisted server-side, enabling access to past fits across sessions. The full process, from submission to result delivery, typically completes within minutes, enabling rapid on-demand triage of transient candidates without waiting for spectroscopic confirmation.

\section{Transient Selection and Data Preparation}
\label{sec:filtering}

\begin{figure}[ht!]
\centering
\includegraphics[width=0.48\textwidth]{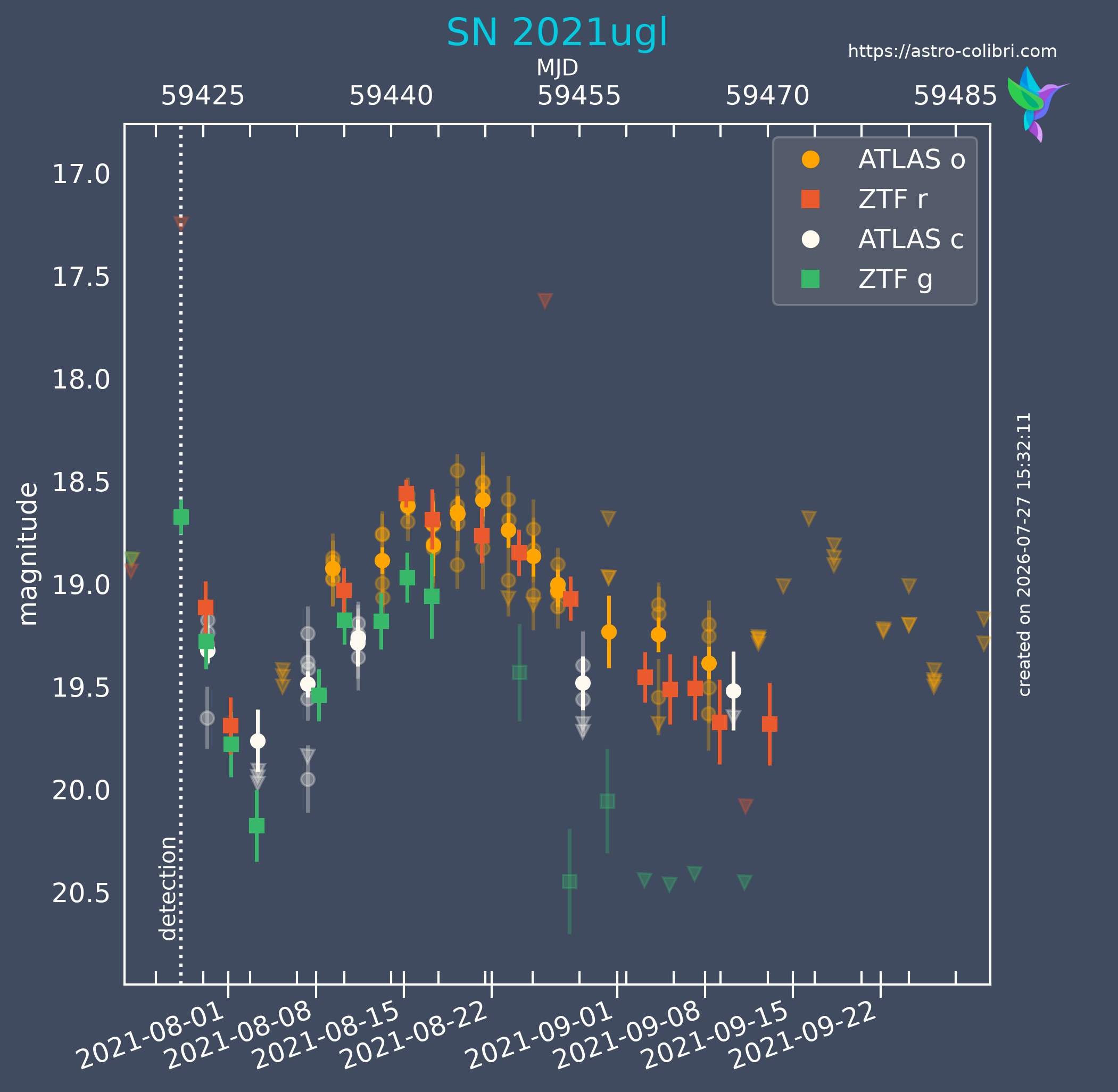}
\caption{Raw archived photometry of SN\,2021ugl as retrieved from the \emph{Astro-COLIBRI} optical light-curve archive, combining ZTF ($g$, $r$) and ATLAS ($c$, $o$) detections and upper limits (ULs, downward triangles) prior to any quality filtering or fitting. The dotted vertical line marks the first reported detection. This is the raw data product passed to the pipeline described above; Section~\ref{sec:results} presents the corresponding NMMA best-fit light curves.}
\label{fig:sn2021ugl-raw-lc}
\end{figure}

In the \texttt{NMMA--Astro-COLIBRI} pipeline, transient selection is driven entirely by the user; only the subsequent classification is automated. \emph{Astro-COLIBRI} continuously ingests alerts from a broad set of brokers and streams (Section~\ref{sec:platform}), including the Fink broker \citep{Moller_2021}, the General Coordinates Network (GCN)\footnote{\scriptsize \url{https://gcn.nasa.gov/}}, and the Transient Name Server (TNS)\footnote{\url{https://www.wis-tns.org/}}. All incoming alerts are displayed in real-time on the web and mobile interfaces where users can browse, filter, and select any transient of interest.

No automated pre-selection or broker-side cut is applied before the fit: any optical transient stored in the \emph{Astro-COLIBRI} archive can be submitted for analysis through the workflow described in Section~\ref{sec:platform}. Any transient carrying one of the three optical-transient tags assigned at ingestion is eligible, provided a photometry file is available. These tags record the TNS status rather than gate the analysis: \texttt{ot} when no TNS classification is yet available, \texttt{ot\_sn} when TNS has classified the object as a supernova of any subtype, and \texttt{ot\_other} when TNS has assigned a non-supernova classification (e.g.\ AGN, nova, cataclysmic variable). This design decouples candidate identification, which is driven by scientific judgment, from quantitative classification, which is handled objectively by the Bayesian framework.

\textbf{Data preparation}---Once a transient is selected, the pipeline retrieves the photometry data aggregated by Astro-COLIBRI and applies per-source quality filtering before passing the data to NMMA. For ZTF, only observations carrying the \texttt{good} quality flag are retained as detections; for ATLAS, only \texttt{average} quality flux-binned measurements are kept. ATLAS forced-photometry epochs individually reach a much shallower detection limit (i.e. a lower signal-to-noise ratio per exposure) than a single ZTF exposure; to reach a stable per-point precision for faint transients, the \emph{Astro-COLIBRI} archive re-bins ATLAS detections (signal-to-noise $\geq 5$) into fixed 90-minute ($0.0625$\,day) windows per filter and observing night, following standard practice for the ATLAS forced-photometry stream \citep{Tonry_2018}. Within each window, the flux is the unweighted mean of the individual-epoch flux densities, $\bar{S} = \tfrac{1}{N}\sum_i S_i$, and its uncertainty follows from standard propagation of independent Gaussian errors, $\sigma_{\bar S} = \tfrac{1}{N}\sqrt{\sum_i \sigma_{S,i}^2}$ (the standard error of the mean). The binned flux is converted to an AB magnitude via $m = -2.5\log_{10}(\bar S/\mu\mathrm{Jy}) + 23.9$, with the magnitude uncertainty from linear error propagation, $\sigma_m = 1.086\,\sigma_{\bar S}/\bar S$ ($1.086 \simeq 2.5/\ln 10$). Windows containing a single detection retain their original single-epoch magnitude and uncertainty unchanged. Non-detections (signal-to-noise $< 5$) are never binned and are kept individually as ULs (see Figure~\ref{fig:sn2021ugl-raw-lc}). ULs are retained by default for both ATLAS and ZTF, and all fits reported in Section~\ref{sec:results} use them: the survival-function term of Equation~\ref{eq:likelihood} constrains the rise time and the early faintness of the transient, tightening the separation between kilonova and shock-cooling models (Section~\ref{subsec:early}).

In the NMMA data format, each detection is encoded as a tuple of observing epoch, filter, magnitude and uncertainty $(t_{\rm ISO},\, f,\, m_{\rm det},\, \sigma_m)$.  The sentinel value $\sigma_m = \infty$ instructs the likelihood engine to interpret the magnitude as an UL and replace the Gaussian detection term with a survival function (Equation~\ref{eq:likelihood}; Section~\ref{subsec:inference}), penalizing any model that predicts a brighter source at that epoch. Photometric filter names are remapped from the \emph{Astro-COLIBRI} convention (e.g.\ \texttt{ztf\_g}) to \texttt{SNCosmo}\footnote{\scriptsize \url{https://github.com/sncosmo/sncosmo}} bandpass names \citep{barbary2016sncosmo, barbary_2025_15019859} that NMMA uses internally (e.g.\ \texttt{ztfg}). The analysis window and photometric bands are defined per run configuration as described in Section~\ref{sec:results}.\\

\begin{figure}[ht!]
\centering
\includegraphics[width=0.48\textwidth]{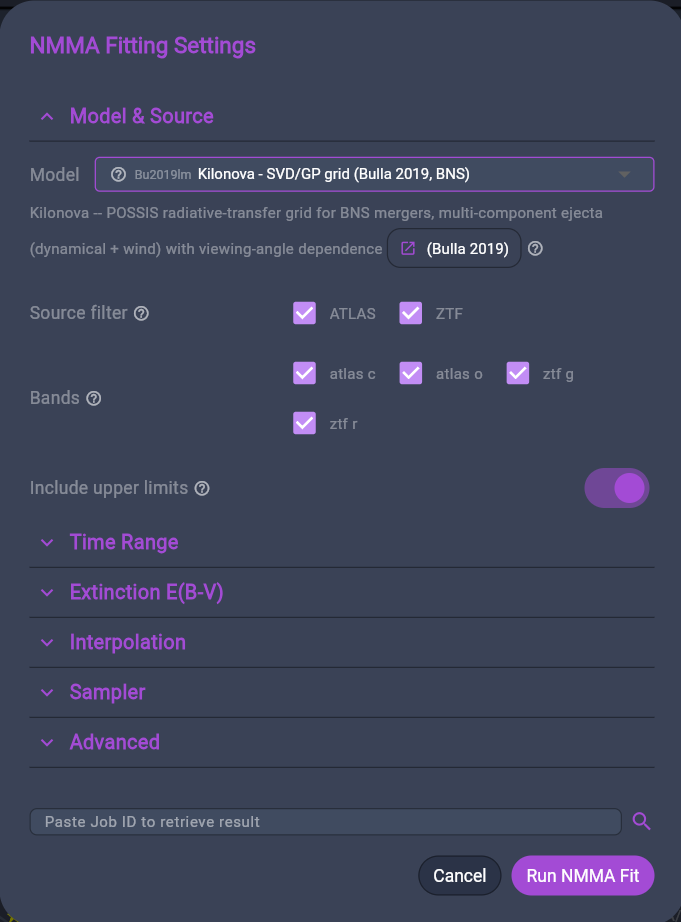}
\caption{The \texttt{NMMA--Astro-COLIBRI} fitting-settings dialog presented to the user before a fit is submitted. The user selects one of the eleven supernova models or five kilonova models (with an inline description and a link to the reference publication), the survey(s) to fit (ZTF/ATLAS/LSST), the analysis time window, and whether to sample the Galactic color excess $E(B\!-\!V)$; sampler and interpolation choices are exposed under collapsible ``Advanced'' sections for users who wish to override the server defaults. Unmodified fields fall back to the defaults described in Section~\ref{sec:nmma}.}
\label{fig:settings-dialog}
\end{figure}

\textbf{Job submission}---Figure~\ref{fig:settings-dialog} shows the settings dialog presented to the user before a fit is submitted. \textbf{Model \& Source}: the user picks one of the eleven supernova models or five kilonova models (Section~\ref{sec:nmma}), shown with an inline description and a link to its reference publication, then restricts the fit to specific surveys (ZTF, ATLAS, LSST) and, within each, specific photometric bands. ULs are retained by default and can be discarded from this dialog. Band selection is applied after the time cut, and any band left without a detection inside the analysis window is excluded from the fit; the set of bands actually used is recorded with each job.
\textbf{Time Range}: the analysis window (\texttt{tmin}/\texttt{tmax}) defaults to the selected survey(s)' own detection coverage and can be overridden via day offsets or a date picker. \textbf{Extinction $E(B\!-\!V)$}: whether the color excess is sampled as a free parameter, its maximum value (pre-filled from the \emph{Astro-COLIBRI} dust-map query, optionally with \citet{Schlafly_2011} recalibration described in Section~\ref{sec:nmma}), and the extinction law (Section~\ref{subsec:inference}) are all user-editable. \textbf{Interpolation} and \textbf{Sampler} expose the SVD interpolation scheme and the nested-sampling algorithm (Section~\ref{subsec:inference}). An \textbf{Advanced} panel exposes the remaining inference settings from Section~\ref{subsec:inference} and Table~\ref{tab:priors}: the number of live points, the time-grid resolution, the systematic error budget $\sigma_{\rm eb}$ (added in quadrature to the photometric measurement uncertainty to account for known imprecision in the light-curve models themselves), the trigger time (timeshift) prior bounds, auto-populated per model from its prior file, and an optional known spectroscopic redshift, which if supplied fixes the luminosity distance to $D_L(z)$ rather than sampling it freely. Any field left unmodified falls back to the server-side default listed in Table~\ref{tab:priors}.

\section{The NMMA Framework and Supernova Models}
\label{sec:nmma}

\subsection{Bayesian inference engine}
\label{subsec:inference}

The NMMA framework \citep{Pang_2023,Rose:2026} performs Bayesian parameter inference on multi-band photometric light curves. For a given model $\mathcal{M}$ with parameters $\vec{\theta}$ and observed data $\mathbf{d}$, the posterior follows from Bayes' equation,
\begin{equation}
    p(\vec{\theta}\,|\,\mathbf{d}, \mathcal{M}) =
    \frac{\mathcal{L}(\mathbf{d}\,|\,\vec{\theta}, \mathcal{M})\;
    \pi(\vec{\theta}\,|\,\mathcal{M})}
    {Z(\mathbf{d}\,|\,\mathcal{M})},
\label{eq:bayes}
\end{equation}
where $\mathcal{L}$, $\pi$, and $Z$ are the likelihood, prior, and Bayesian evidence, respectively. The evidence is obtained by marginalizing over the full parameter space,
\begin{equation}
    Z(\mathbf{d}\,|\,\mathcal{M}) =
    \int \mathcal{L}(\mathbf{d}\,|\,\vec{\theta}, \mathcal{M})\;
    \pi(\vec{\theta}\,|\,\mathcal{M})\; \mathrm{d}\vec{\theta}.
\label{eq:evidence}
\end{equation}

\textbf{Forward model}---For each filter $f$ and epoch $t_n$, the model predicted AB magnitude is
\begin{equation}
    \hat{m}_{n,f}(\vec{\theta}) =
    \mathcal{M}_f\!\left(\phi;\,\vec{\theta}\right)
    + \Delta m
    + 5\log_{10}\!\left(\frac{D_L}{10\,\mathrm{pc}}\right)
    + A_f\!\left(E(B\!-\!V)\right),
\label{eq:forward_model}
\end{equation}
where $\mathcal{M}_f(\phi;\vec{\theta})$ is the template magnitude in observer-frame filter $f$ evaluated at the \emph{source-frame} phase $\phi = (t_n - \tau)/(1+z)$, and $A_f\bigl(E(B\!-\!V)\bigr)$ is the extinction in that filter, computed from the sampled color excess with the law adopted below. Here, $\tau$ is the trigger-time offset (the NMMA \texttt{timeshift}), $D_L$ is the luminosity distance, and the redshift $z = z(D_L)$ is derived from the sampled distance assuming the Planck 2018 cosmology \citep{2020A&A...641A...6P}.

For the template-based models, $\mathcal{M}_f$ is evaluated by \texttt{SNCosmo} which handles the redshift of the spectral template internally (K-correction and $(1+z)$ flux dilution), while cosmological time dilation appears explicitly through the $(1+z)$ factor in the phase argument.

The source-frame peak absolute magnitude of core-collapse supernova templates is first pinned to a fiducial $M_V = -19.35$ (Vega)\footnote{When the template wavelength coverage does not include the Bessell~$V$ filter, the anchor falls back to Bessell~$B$ (Vega) or SDSS~$g$ (AB); the associated color term is absorbed by $\Delta m$.}, so that $\Delta m$ acts as an achromatic offset relative to this reference. For \texttt{salt3}, the amplitude $x_0$ directly parametrizes the apparent flux: the distance-modulus term in Equation~\ref{eq:forward_model} is omitted for this model, $\Delta m$ is not sampled, and $D_L$ enters only through $z$.

For the models implemented natively in NMMA (\texttt{Piro2021} and the kilonova models of Section~\ref{subsec:kn_models}), $\mathcal{M}_f$ is the source-frame absolute magnitude returned by the model itself, $\Delta m \equiv 0$, and the $(1+z)$ flux dilution is applied explicitly as an additional $-2.5\log_{10}(1+z)$ term in place of the internal \texttt{SNCosmo} treatment.

\textbf{Likelihood}---For each filter $f$, the data are partitioned into detections (finite photometric uncertainty $\sigma_{n,f}$) and ULs, i.e. non-detections (encoded with $\sigma_{n,f}\to\infty$). A systematic uncertainty floor $\sigma_{\rm eb} = 0.25$\,mag, supplied through the \texttt{--em-error-budget} argument as a per-filter constant (i.e.\ held fixed, not sampled), is added in quadrature to the photometric uncertainty of detections,
$\Sigma_{n,f} = \sqrt{\sigma_{n,f}^2 + \sigma_{\rm eb}^2}$, accounting for residual template and photometric calibration uncertainties. Its normalization contribution $-\tfrac{1}{2}\ln(2\pi\Sigma_{n,f}^2)$ is identical for all models and cancels exactly in Equation~\ref{eq:bayes_factor}; $\sigma_{\rm eb}$ nevertheless enters the weighting of the residuals and therefore affects each evidence individually. It is held at the same value for
every model, so that no model is advantaged by a different noise budget. The total log-likelihood combines a Gaussian term for the detections and a survival-function term for the ULs:
\begin{equation}
    \begin{split}
        \ln\mathcal{L} = \sum_{f=1}^{F}\Bigg[
            & \sum_{n\in\mathcal{D}_f}
            \left(
              -\frac{1}{2}\frac{\bigl(m_{n,f}-\hat{m}_{n,f}\bigr)^2}{\Sigma_{n,f}^2}
              -\frac{1}{2}\ln\!\bigl(2\pi\,\Sigma_{n,f}^2\bigr)
            \right) \\
            & {} + \sum_{n\in\mathcal{U}_f}
            \ln\!\left[\,1 - \Phi\!\left(
              \frac{m_{n,f} - \hat{m}_{n,f}}{\sigma_{\rm eb}}
            \right)\right] \Bigg],
    \end{split}
    \label{eq:likelihood}
\end{equation}
where $\mathcal{D}_f$ and $\mathcal{U}_f$ are the sets of detections and ULs in filter $f$; $m_{n,f}$ is the measured magnitude (the reported limiting magnitude $m_{n,f}^{\rm lim}$ for an UL); $\hat{m}_{n,f}(\vec{\theta})$ is the model predicted AB magnitude in Equation~\ref{eq:forward_model}; and $\Phi$ is the standard normal cumulative distribution function. The survival-function term penalizes any model that predicts a source brighter than the limiting magnitude at a non-detected epoch; only $\sigma_{\rm eb}$ enters its scale, as non-detected epochs carry no measured photon noise.\footnote{NMMA implements a more general truncated-Gaussian likelihood, in which a per-filter detection limit truncates the Gaussian density of the detection term. We do not use it here: no per-filter detection limit is supplied (\texttt{detection\_limit} $=\infty$ in \texttt{BILBY}; \citealt{Ashton:2018jfp,Romero-Shaw:2020owr}), so the truncation is inactive and the detection term reduces to the standard Gaussian density of Equation~\ref{eq:likelihood}.}

\textbf{Model comparison}---When the user submits fits for two or more models on the same transient (Figure~\ref{fig:model_compare}), the \emph{Astro-COLIBRI} backend computes the log Bayes factor,
\begin{equation}
    \ln\mathcal{B}_{12} =
    \ln Z(\mathbf{d}\,|\,\mathcal{M}_1) -
    \ln Z(\mathbf{d}\,|\,\mathcal{M}_2),
\label{eq:bayes_factor}
\end{equation}
with $\ln\mathcal{B}_{12} > 0$ favoring $\mathcal{M}_1$. Bayesian model comparison via nested-sampling evidences is the core model-selection capability of the NMMA framework \citep{Pang_2023,Rose:2026}; it has been applied, for instance, to the multi-wavelength classification of GRB\,211211A across four astrophysical scenarios \citep{Kunert_2024}, and to the automated daily fitting of ZTF fast-transient candidates \citep{Barna_2024}. We interpret $\ln\mathcal{B}_{12}$ on the Jeffreys scale \citep{Jeffreys1961,Kass_1995}: $|\ln\mathcal{B}_{12}| < 1.1$ is not worth more than a bare mention; $1.1$--$2.3$ constitutes substantial evidence; $2.3$--$3.4$ strong evidence; $3.4$--$4.6$ very strong evidence; and $|\ln\mathcal{B}_{12}| > 4.6$ decisive evidence. Because $Z$ integrates the likelihood over the full prior volume (Equation~\ref{eq:evidence}), models with additional free parameters that do not improve the fit are penalized by an Occam factor \citep{MacKa_1992}, so Equation~\ref{eq:bayes_factor} embodies a complexity-penalizing comparison rather than a simple goodness-of-fit ratio. In what follows, we report $\ln\mathcal{B} = \ln Z_{\mathrm{ref}} - \ln Z_i$, where $\mathcal{M}_{\mathrm{ref}}$ is the highest-evidence model of each run, so that positive values quantify how strongly model $\mathcal{M}_i$ is disfavored with respect to the best-supported one.

\textbf{Goodness-of-fit diagnostic}---After each nested-sampling run, the maximum \textit{a posteriori} (MAP) parameter vector is identified from the posterior samples and a per-filter reduced chi-squared $\chi^2_{{\rm red},f}$ is computed as a residual diagnostic:
\begin{equation}
  \chi^2_{{\rm red},f} =
  \frac{1}{N_f^{\rm det}}
  \sum_{n\in\mathcal{D}_f}
  \frac{\bigl(m_{n,f} - \hat{m}_{n,f}(\vec{\theta}_{\rm MAP})\bigr)^2}
       {\sigma_{n,f}^2 + \sigma_{\rm eb}^2},
\label{eq:chi2}
\end{equation}
where $N_f^{\rm det}$ is the number of detections in filter $f$. ULs do not enter Equation~\ref{eq:chi2}; they contribute to the evidence through the survival-function term in Equation~\ref{eq:likelihood}.

\textbf{Sampling}---Nested sampling \citep{Skilling:2006gxv,Veitch:2009hd} is performed with the \texttt{BILBY} library \citep{Ashton:2018jfp,Romero-Shaw:2020owr}, which supports several samplers such as \texttt{DYNESTY} \citep{2020MNRAS.493.3132S,sergey_koposov_2022_6456387}; our pipeline uses \texttt{MultiNest} \citep{Feroz_2009} via \texttt{PyMultiNest} \citep{Buchner2014} with 2048 live points by default.

\textbf{Priors}---All models share three common free parameters: the luminosity distance $D_L$, the trigger-time offset $\tau$, and the Galactic color excess $E(B\!-\!V)$ (Table~\ref{tab:priors}). Additional model-specific parameters are listed in Table~\ref{tab:priors} for supernova and Table~\ref{tab:kn_priors} for kilonova models.

\textbf{Extinction}---Two extinction laws are implemented and selectable in the pipeline: the \citet{Gordon_2023} Milky-Way average curve with $R_V = 3.1$, evaluated in the observer frame, and the \citet{Pei_1992} SMC (Small Magellanic Cloud) curve with $R_V = 2.93$, evaluated at the transient redshift. All fits in this work use the former, the physically consistent choice when $E(B\!-\!V)$ is bounded by a Galactic dust map; at the reddening levels relevant here, the two options agree to within $0.02$~mag in the ZTF and ATLAS bands, well below $\sigma_{\rm eb}$. 

The color excess is a free parameter with a uniform prior over $[0,\,E(B\!-\!V)_{\rm max}]$, where $E(B\!-\!V)_{\rm max}$ is the Galactic line-of-sight reddening at the target coordinates, obtained by the \emph{Astro-COLIBRI} back-end from the \citet{Schlegel_1998} dust map via \texttt{dustmaps} \citep{Green_2018}.

The settings dialog offers an optional toggle for the $0.86$ rescaling of \citet{Schlafly_2011}, which the user may activate to correct the known $\sim$14\% overestimate of the original map calibration; this rescaling is the calibration consistent with the $R_V = 3.1$ Milky-Way law adopted here, whereas NMMA itself queries the map without recalibration. All fits reported in this work activate this option, and for SN\,2021ugl this gives $E(B\!-\!V)_{\rm max} = 0.057$~mag.

\begin{table*}[ht!]
\centering
\caption{Prior distributions used in all \texttt{NMMA--Astro-COLIBRI} fits. The three shared parameters ($D_L$, $\tau$, $E(B\!-\!V)$) are sampled for every model; model-specific parameters are sampled only for the models listed in the last column.}
\label{tab:priors}

\begin{tabular}{lllrrl}
\toprule
\hline
Parameter & Symbol & Prior & $\theta_{\min}$ & $\theta_{\max}$ & Applies to \\
\midrule

\hline
\addlinespace[4pt]
\multicolumn{6}{l}{\textbf{\textit{Shared parameters (all models)}}} \\
\addlinespace[4pt]
Luminosity distance (Mpc)        & $D_L$        & Uniform        & $10^{-3}\, ^{a}$    & $7000$        & All$^{b}$ \\
Trigger-time offset (days)       & $\tau$        & Uniform        & $-30$   & $+30$          & All\\
Galactic colour excess (mag) & $E(B\!-\!V)$ & Uniform & $0$ & $E(B\!-\!V)_{\rm max}$ & All \\
\hline

\addlinespace[8pt]
\multicolumn{6}{l}{\textbf{\textit{Model-specific parameters}}} \\
\addlinespace[4pt]
Template magnitude boost (mag)   & $\Delta m$        & Uniform & $-5$   & $+5$   & \texttt{SNCosmo} (except \texttt{salt3}) \\
\addlinespace[2pt]
SALT3 flux normalization         & $x_0$             & Uniform & $0$    & $10$   & \texttt{salt3} \\
SALT3 light-curve shape          & $x_1$             & Uniform & $-5$   & $+5$   & \texttt{salt3} \\
SALT3 color (mag)               & $c$               & Uniform & $-0.5$ & $+2.0$ & \texttt{salt3} \\
\addlinespace[2pt]
Envelope mass ($\log_{10}[M_e/M_\odot]$)        & $\log_{10}M_e$ & Uniform & $-1.5$ & $-0.5$ & \texttt{Piro2021} \\
Envelope radius ($\log_{10}[\mathrm{cm}]$)      & $\log_{10}R_e$ & Uniform & $11.3$ & $14.3$ & \texttt{Piro2021} \\
Extended envelope energy ($\log_{10}[\mathrm{erg}]$) & $\log_{10}E_e$ & Uniform & $48.4$ & $50.8$ & \texttt{Piro2021} \\
\bottomrule
\end{tabular}

\vspace{2pt}
\begin{minipage}{\textwidth}
\footnotesize
$^{a}$ A lower bound of $10^{-3}$~Mpc is imposed on $D_L$, so that \texttt{astropy}'s $D_L\!\to\!z$ inversion remains numerically well defined.

$^{b}$ \texttt{Piro2021} caps $D_L$ at $1000$~Mpc rather than $7000$~Mpc.

\end{minipage}
\end{table*}

\textbf{Brightness degeneracy}---Because the photometry constrains only the combination $\mu(D_L) + \Delta m$, where $\mu = 5\log_{10}(D_L/10\,\mathrm{pc})$, the luminosity distance and the magnitude boost are degenerate: their marginal posteriors trace the prior along the degeneracy direction and are not reported as physical measurements (Equation~\ref{eq:forward_model}). This choice is deliberate for \emph{classification}: all anchored templates share identical priors on $(D_L, \Delta m, \tau, E(B\!-\!V))$, so the associated Occam factors are common to all models and cancel in the log Bayes factors of Equation~\ref{eq:bayes_factor}. When a spectroscopic redshift is available, the service can instead fix $D_L$ at the corresponding distance and sample only $\Delta m$, restoring a physical interpretation of the brightness offset.

\subsection{Supernova models}
\label{subsec:sn_models}

The current \texttt{NMMA--Astro-COLIBRI} service implements eleven models spanning the principal supernova subclasses: thermonuclear (Type\,Ia), hydrogen-rich core-collapse (Types\,IIP, IIL, IIn), the transitional Type\,IIb, stripped-envelope (Type\,Ib/c), and an analytical shock-cooling model. Ten template-based models are evaluated through the \texttt{SNCosmo} library \citep{barbary2016sncosmo}; the shock-cooling model (\texttt{Piro2021}) is implemented analytically in NMMA. All \texttt{SNCosmo}-based models except \texttt{salt3} share the same prior on $D_L$, $\tau$, $E(B\!-\!V)$, and $\Delta m$ (Table~\ref{tab:priors}), keeping the total number of sampled parameters, $N_\theta$, at 4 for nine of the eleven models. \texttt{salt3} and \texttt{Piro2021} instead each sample three model-specific parameters in place of $\Delta m$, giving $N_\theta = 6$.

\textbf{Type\,Ia -- SALT3}---The \texttt{salt3} model \citep{Kenworthy_2021} is a spectral adaptive light-curve template for Type\,Ia \acp{SN}, extending \texttt{salt2} \citep{Guy:2007dv} over 2000--11000~\AA. Its parameters are a flux normalization $x_0$, a shape parameter $x_1$, and a color term $c$. Calibrated on a large low-redshift sample, it is the standard tool for cosmological \ac{SNIa} analysis.

\textbf{Type\,Ia -- Nugent}---The \texttt{nugent-sn1a} template \citep{Nugent_2002} provides an alternative \ac{SNIa} spectral energy distribution through \texttt{SNCosmo}, with a simpler parameterization than \texttt{salt3}; it is retained as a complementary check on the Type\,Ia classification.

\textbf{Type\,IIP}---The \texttt{nugent-sn2p} template \citep{Gilliland_1999} models hydrogen-rich Type\,IIP \acp{SN}, whose extended plateau ($\sim$80--100~days) arises from recombination of a massive hydrogen envelope.

\textbf{Type\,IIL}---The \texttt{nugent-sn2l} template \citep{Gilliland_1999} models Type\,IIL \acp{SN}, which show a linear post-peak decline rather than a plateau, indicating a less massive hydrogen envelope than Type\,IIP events.

\textbf{Type\,IIn}---The \texttt{nugent-sn2n} template \citep{Gilliland_1999} models Type\,IIn \acp{SN}, whose spectra are dominated by narrow, multi-component Balmer emission (notably H$\alpha$) produced by interaction of the ejecta with dense circumstellar material.

\textbf{Type\,IIb -- SN\,1993J}---The \texttt{v19-1993j-corr} template \citep{Vincenzi_2019} is the host-extinction corrected, daily-sampled spectrophotometric time series built for SN\,1993J, the prototypical Type\,IIb supernova \citep{Filippenko_1993}. Type\,IIb events are transitional: they show early hydrogen lines that fade as the thin residual envelope is shed, eventually resembling a Type\,Ib spectrum. The template is fully data-driven, reconstructed from the multi-band photometry and sparse spectroscopy of SN\,1993J through two-dimensional Gaussian-process interpolation with no assumed parametric \ac{SED}, and extended into the near-ultraviolet. Critically for early classification, the shock-breakout bump is modelled explicitly in the light-curve fit \citep{Vincenzi_2019}, so the template reproduces the characteristic double-peaked Type\,IIb morphology: an early luminosity excess from post-shock-breakout cooling of the extended, low-mass hydrogen envelope, followed by the radioactively ($^{56}$Ni/$^{56}$Co) powered second maximum near $\sim$20~days post-explosion \citep{Filippenko_1993}. Because a single template encodes both phases self-consistently, no separate analytical shock-cooling component is required. The \texttt{-corr} suffix denotes host-galaxy dust correction, yielding intrinsic broad-band colors. Its only free parameter beyond the shared prior is $\Delta m$.

\textbf{Type\,Ib/c}---The \texttt{nugent-sn1bc} template \citep{LeNu2005} models normal stripped-envelope \acp{SNIbc}, massive stars that have shed their hydrogen (Ib) and, for Ic, helium envelopes, through \texttt{SNCosmo}. Its primary free parameter is the absolute-magnitude normalization.

\textbf{Hypernova -- high-velocity Ib/c}---The \texttt{nugent-hyper} template \citep{LeNu2005}, built from the same study, represents the high-velocity Type\,Ib/c events associated with long \acp{GRB} (e.g., GRB\,980425/SN\,1998bw). It is constructed from the spectra of SN\,1998bw, whose temporal evolution matches the other high-velocity events SN\,1997ef and SN\,2002ap; its peak is normalized to $M_V \approx -17.4$, characteristic of that class. \citet{LeNu2005} explicitly contrast these high-velocity supernovae with normal, lower-velocity stripped-envelope events such as SN\,1993J (IIb) and SN\,1994I (Ic), which decline more rapidly, the same velocity distinction that photometric classification of SN\,2021ugl relies on. As for all \texttt{SNCosmo} templates in this work, the amplitude is re-anchored to $M_V=-19.35$~mag (Vega) before fitting (Section~\ref{subsec:inference}); the native normalisation $M_V\approx-17.4$ of \texttt{nugent-hyper} is therefore absorbed into the fitted $\Delta m\approx+1.9$~mag relative to the fiducial reference.
 
\textbf{Type\,Ib}---The \texttt{v19-2008d-corr} template \citep{Vincenzi_2019} models Type\,Ib supernovae, built from the well-studied prototype SN\,2008D. Type\,Ib events have shed their hydrogen envelope but retain helium, shown through prominent He\,\textsc{i} lines. Its only free parameter beyond the shared prior is $\Delta m$.

\textbf{Type\,Ic}---The \texttt{v19-1994i-corr} template \citep{Vincenzi_2019} models Type\,Ic supernovae, built from the prototypical SN\,1994I. Type\,Ic events have lost both their hydrogen and helium envelopes, lacking both H (Balmer) and He\,\textsc{i} lines. Its only free parameter beyond the shared prior is $\Delta m$.

\textbf{Shock cooling (Piro2021)}---The \texttt{Piro2021} model \citep{PiHa2021} describes early-time emission from post-shock-breakout cooling of extended material surrounding the progenitor, with three parameters: the envelope mass $M_e$, radius $R_e$, and energy $E_e$. It fades on day timescales like a genuine kilonova, making it the library's most astrophysically critical model for kilonova discrimination.

\begin{figure}[ht!]
\centering
\includegraphics[width=0.48\textwidth]{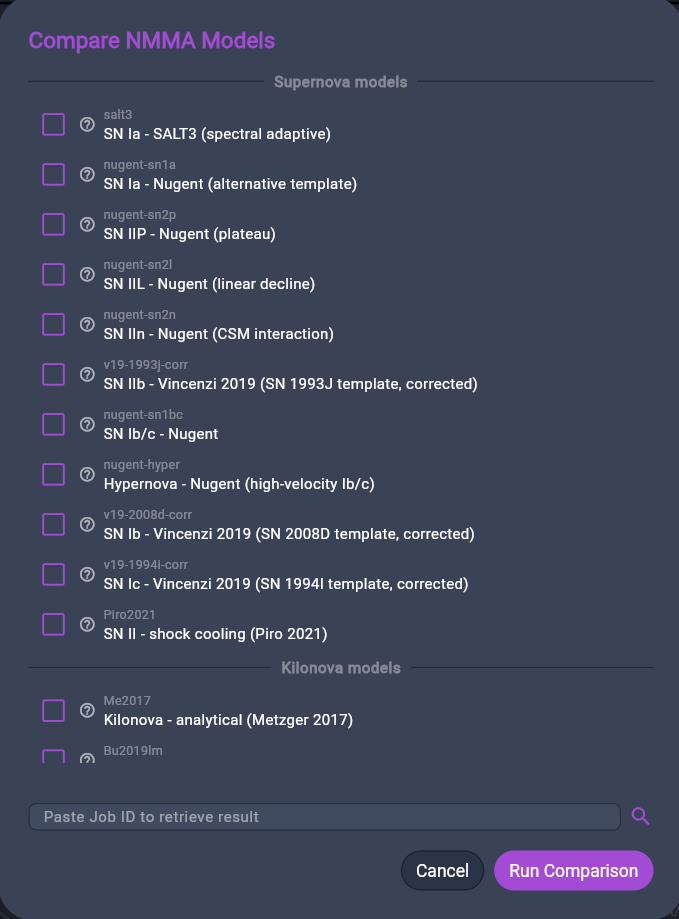}
\caption{The \texttt{NMMA--Astro-COLIBRI} model-comparison dialog, listing the eleven supernova and five kilonova models available for a joint fit (Sections~\ref{subsec:sn_models}--\ref{subsec:kn_models}), each identified by its model key and a short physical description. The user checks two or more models and submits them as a batch via \textbf{Run Comparison}; once all fits complete, the back-end ranks the selected models by their log Bayes factor (Equation~\ref{eq:bayes_factor}). A job-ID field allows a previously submitted comparison to be retrieved without resubmission.}
\label{fig:model_compare}
\end{figure}

Table~\ref{tab:models} summarizes the eleven \ac{SN} models currently deployed, together with their physical class, implementation route, additional free parameters, and total dimensionality $N_\theta$. The \emph{Model key} column lists the native NMMA/SNCosmo model identifier, reused unchanged by the \emph{Astro-COLIBRI} model-selection interface (Figure~\ref{fig:model_compare}) and used throughout the pipeline to label the corresponding fit results and light-curve/corner plots.

\begin{table*}[ht!]
\centering
\caption{Supernova models implemented in the \texttt{NMMA--Astro-COLIBRI} service. 
All models share the three parameters ($D_L$, $\tau$, $E(B\!-\!V)$) listed in the 
upper block of Table~\ref{tab:priors}; the column ``Additional params.''\ lists only 
the model-specific parameters. $N_\theta$ is the total number of sampled parameters.
$\Delta m$ denotes the template magnitude boost.}
\label{tab:models}
\begin{tabular}{lllll}
\hline\hline
Model key & SN class & Implementation & Additional params. & $N_\theta$ \\
\hline
\texttt{salt3}           & Type\,Ia             & SNCosmo & $x_0,\,x_1,\,c$    & 6 \\
\texttt{nugent-sn1a}     & Type\,Ia             & SNCosmo & $\Delta m$         & 4 \\
\texttt{nugent-sn2p}     & Type\,IIP            & SNCosmo & $\Delta m$         & 4 \\
\texttt{nugent-sn2l}     & Type\,IIL            & SNCosmo & $\Delta m$         & 4 \\
\texttt{nugent-sn2n}     & Type\,IIn            & SNCosmo & $\Delta m$         & 4 \\
\texttt{v19-1993j-corr}  & Type\,IIb            & SNCosmo & $\Delta m$         & 4 \\
\texttt{v19-2008d-corr}  & Type\,Ib             & SNCosmo & $\Delta m$         & 4 \\
\texttt{v19-1994i-corr}  & Type\,Ic             & SNCosmo & $\Delta m$         & 4 \\
\texttt{nugent-sn1bc}    & Type\,Ib/c           & SNCosmo & $\Delta m$         & 4 \\
\texttt{nugent-hyper}    & Type\,Ib/c (hypern.) & SNCosmo & $\Delta m$         & 4 \\
\texttt{Piro2021}        & Shock cooling        & NMMA    & $M_e,\,R_e,\,E_e$ & 6 \\
\hline\hline
\end{tabular}
\end{table*}

\subsection{Kilonova models}
\label{subsec:kn_models}
In addition to the eleven supernova templates of Section~\ref{subsec:sn_models}, the \texttt{NMMA--Astro-COLIBRI} picker exposes a library of five kilonova models. Because kilonovae are intrinsically rare and arise exclusively from compact binary mergers involving at least one neutron star (BNS and NSBH systems), they are astrophysically expected only in spatial and temporal coincidence with a gravitational-wave trigger from the LIGO--Virgo--KAGRA network; outside of an active observing run these models chiefly serve as an adversarial benchmark against the shock-cooling supernovae that mimic their early-time evolution, as applied to SN\,2021ugl in Section~\ref{sec:results}.

The library comprises: the grid-based i) \texttt{Bu2019lm} (BNS) and ii) \texttt{Bu2019nsbh} (NSBH) templates built with the radiative-transfer code \textsc{Possis} \citep{Bulla_2019, DiCo2020, Bulla:2022mwo}, parameterized by the dynamical and disk-wind ejecta masses ($M_{\rm ej}^{\rm dyn}$, $M_{\rm ej}^{\rm wind}$), the half-opening angle of the lanthanide-rich component $\Phi$, and the viewing angle $\theta_{\rm obs}$; iii) its successor \texttt{Bu2023Ye} \citep{Anand_2023}, which replaces the bivalent $\Phi$ prescription for the dynamical ejecta with a continuous, numerical-relativity-guided angular electron-fraction profile $Y_e^{\rm dyn}(\theta) = a\cos^2\theta + b$, using the dynamical-ejecta mass and mean velocity ($M_{\rm ej}^{\rm dyn}$, $ v_{\rm ej}^{\rm dyn}$), the mean dynamical electron fraction $Y_e^{\rm dyn}$, the wind-ejecta mass and mean velocity ($M_{\rm ej}^{\rm wind}$, $ v_{\rm ej}^{\rm wind}$), and the viewing angle as free parameters; iv) the semi-analytic \texttt{Ka2017} model \citep{KaMe2017}, controlled by the ejecta mass $M_{\rm ej}$, expansion velocity $v_{\rm ej}$, and lanthanide fraction $X_{\rm lan}$; v) the single-component analytic \texttt{Me2017} model \citep{Metzger_2017, Metzger_2019}, based on a semi-analytic light-curve solution for ejecta with a power-law mass-velocity distribution $M_v = M\,(v/v_0)^{-\beta}$ and a gray opacity $\kappa$ set by the ejecta's lanthanide content, with free parameters $M$, $v_0$, $\beta$, and $\kappa$.

Table~\ref{tab:kn_models} summarizes the five kilonova models available in the picker, together with their ejecta structure, implementation route, additional free parameters, and total dimensionality $N_\theta$.

\begin{table*}[ht!]
\centering
\caption{Kilonova models implemented in the \texttt{NMMA--Astro-COLIBRI} service. All models share the three parameters ($D_L$, $\tau$, $E(B\!-\!V)$) listed in the upper block of Table~\ref{tab:kn_priors}; the column ``Additional params.''\ lists only the model-specific parameters. $N_\theta$ is the total number of sampled
parameters.}
\label{tab:kn_models}
\begin{tabular}{lllll}
\hline\hline
Model key & Components & Implementation & Additional params. & $N_\theta$ \\
\hline
\texttt{Me2017}    & One (analytic)         & NMMA (analytic)  & $M,\,v_0,\,\beta,\,\kappa$                                                & 7  \\
\texttt{Bu2019lm}  & Two (BNS)                 & NMMA (SVD grid)  & $M_{\rm ej}^{\rm dyn},\,M_{\rm ej}^{\rm wind},\,\Phi,\,\iota$             & 7  \\
\texttt{Bu2019nsbh}& Two (NSBH)                & NMMA (SVD grid)  & $M_{\rm ej}^{\rm dyn},\,M_{\rm ej}^{\rm wind},\,\iota$                    & 6  \\
\texttt{Ka2017}    & One (1D grid)          & NMMA (SVD grid)  & $M_{\rm ej},\,v_{\rm ej},\,X_{\rm lan}$                                   & 6  \\
\texttt{Bu2023Ye}  & Two (BNS, $Y_e$-parameterized) & NMMA (SVD grid) & $M_{\rm ej}^{\rm dyn},\, v_{\rm ej}^{\rm dyn},\, Y_e^{\rm dyn},\,M_{\rm ej}^{\rm wind},\, v_{\rm ej}^{\rm wind},\, Y_e^{\rm wind},\,\iota$ & 10 \\
\hline\hline
\end{tabular}
\end{table*}

\begin{table*}[ht!]
\centering
\caption{Prior distributions for the five kilonova models implemented in the \texttt{NMMA--Astro-COLIBRI} picker (Section~\ref{subsec:kn_models}). The shared luminosity-distance and trigger-time priors are substantially narrower than the supernova priors of Table~\ref{tab:priors}, reflecting the much closer distances and faster time evolution of kilonovae.}
\label{tab:kn_priors}
\begin{tabular}{lllrrl}
\toprule
\hline
Parameter & Symbol & Prior & $\theta_{\min}$ & $\theta_{\max}$ & Applies to \\
\midrule
\hline
\addlinespace[4pt]
\multicolumn{6}{l}{\textbf{\textit{Shared parameters (all kilonova models)}}} \\
\addlinespace[4pt]
Luminosity distance (Mpc)    & $D_L$ & Uniform & $0$    & $200$ & All KN \\
Trigger-time offset (days)   & $\tau$ & Uniform & $-2$  & $+1$  & All KN \\
Galactic colour excess (mag) & $E(B\!-\!V)$ & Uniform & $0$ & $E(B\!-\!V)_{\rm max}$ & All KN \\
\hline

\addlinespace[8pt]
\multicolumn{6}{l}{\textbf{\textit{Model-specific parameters}}} \\
\addlinespace[4pt]
Ejecta mass ($\log_{10}[M_\odot]$)      & $\log_{10}M_{\rm ej}$      & Uniform & $-3.0$  & $-0.5$  & \texttt{Me2017} \\
Ejecta velocity ($\log_{10}[c]$)        & $\log_{10}v_{\rm ej}$      & Uniform & $-2.0$  & $-0.5$  & \texttt{Me2017} \\
Mass-shell profile                      & $\beta$                    & Uniform & $1$     & $5$     & \texttt{Me2017} \\
Opacity normalization ($\log_{10}$)     & $\log_{10}\kappa$        & Uniform & $-1$    & $2$     & \texttt{Me2017} \\
\addlinespace[2pt]
Dynamical ejecta mass ($\log_{10}[M_\odot]$) & $\log_{10}M^{\rm dyn}_{\rm ej}$ & Uniform & $-3.0$ & $-1.0$ & \texttt{Bu2019lm}, \texttt{Bu2019nsbh} \\
Wind ejecta mass ($\log_{10}[M_\odot]$)      & $\log_{10}M^{\rm wind}_{\rm ej}$ & Uniform & $-3.0$ & $-0.5$ & \texttt{Bu2019lm}, \texttt{Bu2019nsbh} \\
Half-opening angle (deg)                     & $\Phi$                     & Uniform & $15$   & $75$   & \texttt{Bu2019lm}$^{a}$ \\
Viewing angle (rad)                          & $\iota$                    & Sine    & $0$    & $\pi/2$ & \texttt{Bu2019lm}, \texttt{Bu2019nsbh}, \texttt{Bu2023Ye} \\
\addlinespace[2pt]
Ejecta mass ($\log_{10}[M_\odot]$)      & $\log_{10}M_{\rm ej}$      & Uniform & $-3.0$  & $-1.0$  & \texttt{Ka2017} \\
Ejecta velocity ($\log_{10}[c]$)        & $\log_{10}v_{\rm ej}$      & Uniform & $-1.52$ & $-0.53$ & \texttt{Ka2017} \\
Lanthanide fraction ($\log_{10}$)       & $\log_{10}X_{\rm lan}$     & Uniform & $-9$    & $-1$    & \texttt{Ka2017} \\
\addlinespace[2pt]
Dynamical ejecta mass ($\log_{10}[M_\odot]$) & $\log_{10}M^{\rm dyn}_{\rm ej}$ & Uniform & $-3.0$  & $-1.7$  & \texttt{Bu2023Ye} \\
Dynamical ejecta velocity ($c$)              & $v^{\rm dyn}_{\rm ej}$      & Uniform & $0.12$  & $0.25$  & \texttt{Bu2023Ye} \\
Dynamical electron fraction                  & $Y_{e,\rm dyn}$             & Uniform & $0.15$  & $0.30$  & \texttt{Bu2023Ye} \\
Wind ejecta mass ($\log_{10}[M_\odot]$)      & $\log_{10}M^{\rm wind}_{\rm ej}$ & Uniform & $-2.0$  & $-0.89$ & \texttt{Bu2023Ye} \\
Wind ejecta velocity ($c$)                   & $v^{\rm wind}_{\rm ej}$     & Uniform & $0.03$  & $0.15$  & \texttt{Bu2023Ye} \\
Wind electron fraction                       & $Y_{e,\rm wind}$            & Uniform & $0.20$  & $0.40$  & \texttt{Bu2023Ye} \\
\bottomrule
\end{tabular}

\vspace{2pt}
\begin{minipage}{\textwidth}
\footnotesize
$^{a}$ For \texttt{Bu2019nsbh} the half-opening angle is fixed to $\Phi = 30^\circ$ in the underlying POSSIS grid and is not sampled.
\end{minipage}
\end{table*}


\subsection{GRB afterglow model}
\label{subsec:grb_model}
Although the present service does not yet expose GRB afterglow fitting, the NMMA framework also implements the \texttt{TrPi2018} GRB afterglow model through \texttt{afterglowpy}\footnote{\scriptsize \url{https://github.com/geoffryan/afterglowpy}} \citep{TrPi2018, RyEe2020}. This model computes forward-shock synchrotron emission from a structured relativistic jet as a function of jet and observer geometry. Its key parameters are the isotropic kinetic energy $E_{\rm K,iso}$, the jet collimation angle $\theta_c$, the viewing angle $\theta_{\rm obs}$, the circumburst density $n$, the electron spectral index $p$, and the energy fractions $\epsilon_e$ and $\epsilon_B$ imparted to electrons and to the magnetic field, respectively. While not among the eleven \ac{SN} subtypes offered in the current \texttt{NMMA--Astro-COLIBRI} interface, the GRB afterglow model remains available in the underlying framework and will be incorporated in a future release of the service.

\section{Results}
\label{sec:results}

\subsection{SN\,2021ugl: a Type\,IIb supernova mistaken for a kilonova candidate}
\label{subsec:sn2021ugl}
 
SN\,2021ugl (ZTF21abotose) was discovered by ZTF on 2021 July~28 \citep{2021TNSTR2582....1M} and spectroscopically confirmed as a Type\,IIb supernova at $z = 0.0412$ ($D_L \simeq 188$~Mpc for the cosmology adopted in Section~\ref{subsec:inference}) by \citet{Ridley_2021} on 2021 August~13, sixteen days after discovery. Its photometric evolution is characteristic of the archetypal Type\,IIb SN\,1993J: a rapid early rise and a first maximum powered by post-shock-breakout cooling of the extended hydrogen envelope, followed by a radioactively powered secondary peak. This double-humped morphology initially led automated alert pipelines to flag the event as a kilonova candidate \citep{Aivazyan_2022, Barna_2024}, making it a controlled, well-characterized test case for the \texttt{NMMA--Astro-COLIBRI} classification service.
 
We retrieved the combined ZTF ($g$, $r$) and ATLAS ($o$) photometry from the \emph{Astro-COLIBRI} archive, covering 47 days of observations from 28 July to 13 September 2021.

\subsection{Analysis configurations}
\label{subsec:configs}

We perform three complementary analyses. An \emph{early-time} configuration restricts the data to the first $\sim\!6$~days after the first ZTF detection, using the two ZTF bands ($g$, $r$). Crucially, this first epoch is a \emph{detection}, not an UL: it already samples the \emph{declining} phase of the shock-cooling emission, so the cooling tail is constrained from $t=0$ rather than merely bracketed by non-detections. This emulates the information available in real-time when the service would first be invoked, on 2021~August~3, ten days before the spectroscopic classification of \citet{Ridley_2021}. The \emph{full-baseline} configuration then fits all three bands (\textit{ztfg}, \textit{ztfr}, \textit{atlas o}) over the complete 47-day window, including the reported ULs through the survival-function term of Equation~\ref{eq:likelihood}; it constitutes our reference classification. A \emph{detections-only} variant repeats this fit with the ULs removed, isolating their contribution to the model ranking. All analyses share the priors of Table~\ref{tab:priors} and Table~\ref{tab:kn_priors}, the systematic floor $\sigma_{\rm eb} = 0.25$~mag, and the extinction treatment of Section~\ref{sec:nmma}; following the brightness-degeneracy discussion, the marginal posteriors of $D_L$ and $\Delta m$ are not interpreted as physical measurements.

\subsubsection{Early-time classification}
\label{subsec:early}

\begin{figure*}[ht!]
\centering
\includegraphics[width=\textwidth]{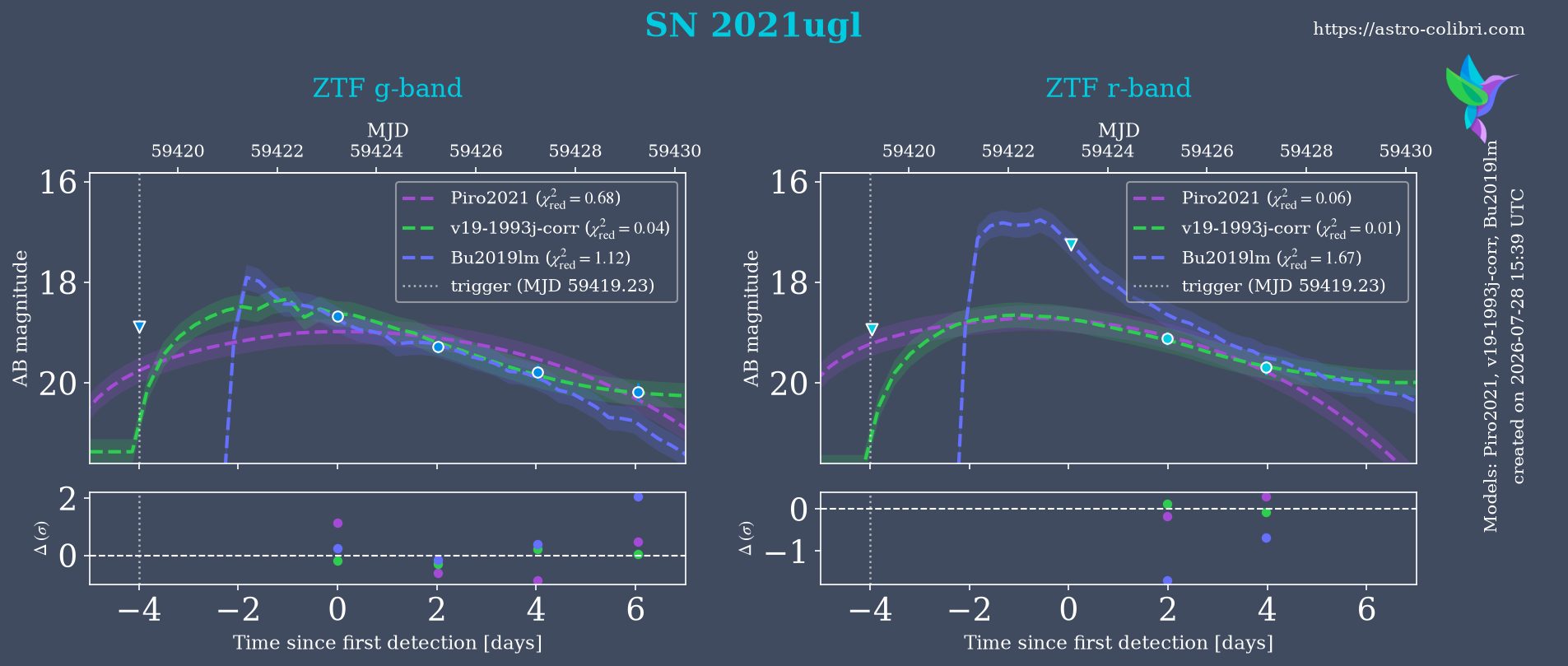}\\[10pt]
\includegraphics[width=\textwidth]{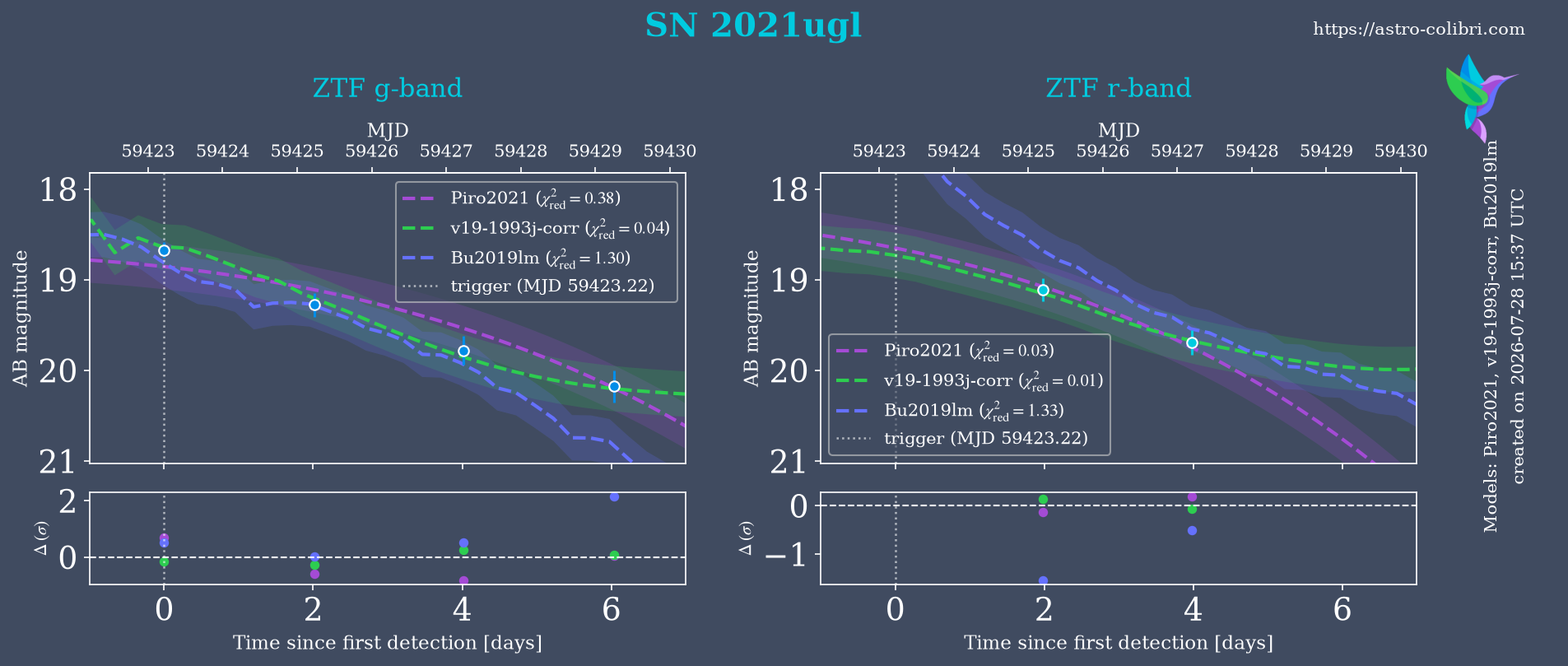}
\caption{Early-time classification of SN\,2021ugl restricted to the first $\sim\!6$~days of ZTF $g$- and $r$-band photometry (available 3~August~2021, ten days before spectroscopic confirmation). \textbf{Top}: fit including pre-trigger and same-night ULs. \textbf{Bottom}: detections-only fit. In both configurations, the empirical Type\,IIb template \texttt{v19-1993j-corr} (green) and the analytic shock-cooling model \texttt{Piro2021} (light purple) track the declining shock-cooling phase sampled from the first detection, whereas the kilonova template \texttt{Bu2019lm} (light blue) over-predicts the early-time brightness and is decisively rejected in both cases ($\ln\mathcal{B}=9.12$ detections-only, $9.83$ with ULs; Table~\ref{tab:early}). Bottom sub-panels show the per-epoch residuals in units of $\sigma$.}
\label{fig:lc_early}
\end{figure*}

Restricting the fit to the first $6$~days (Table~\ref{tab:early}, Figure~\ref{fig:lc_early}) tests whether the service reaches the correct classification in real-time, before any spectrum is available. Because the first epoch is a detection sampling the \emph{declining} shock-cooling phase, single-peaked templates are structurally excluded: over this window \texttt{nugent-hyper} \emph{rises} towards its single maximum and cannot reproduce a decline from $t=0$, whereas both the empirical Type\,IIb template \texttt{v19-1993j-corr} and the analytic shock-cooling model \texttt{Piro2021} track the observed fading in both ZTF bands. We overlay the two as independent fits rather than summing them into a composite: \texttt{v19-1993j-corr} already contains the cooling peak, so adding \texttt{Piro2021} on top would double-count the same emission. Their agreement on the early decline confirms that the first phase is shock-cooling of an extended envelope, i.e.\ a stripped-envelope progenitor. Here \texttt{v19-1993j-corr} is the highest-evidence model, and the log Bayes factors of Table~\ref{tab:early} show that this is not merely a good fit but a decisive classification in both the with-ULs and detections-only configurations: NMMA rejects the kilonova and single-peaked supernova hypotheses ten days before the spectroscopic confirmation.

\textbf{Kilonova}---The kilonova template \texttt{Bu2019lm} \citep{Bulla_2019, DiCo2020, Bulla:2022mwo} is decisively rejected in the early-time window ($\ln\mathcal{B}=9.12$ detections-only, $9.83$ with ULs; Table~\ref{tab:early}), despite a moderately degraded but not catastrophic per-band fit ($\chi^2_{\rm red}\simeq1.1$--$1.7$ vs.\ $\chi^2_{\rm red}=0.04$ ($g$) and $0.01$ ($r$) for \texttt{v19-1993j-corr}). This reflects two compounding effects. First, a rapid rise and decline are required to reproduce the smoother shock-cooling decline, leading to a genuine morphological mismatch as they over-predict the early-time brightness on a timescale of $\lesssim1$~day. Second, an Occam penalty is intrinsic to the obtained evidence: \texttt{Bu2019lm} carries seven free parameters (viewing angle, half-opening angle, ejecta mass for two components, distance, timeshift, and reddening) versus four for the \ac{SN} templates. $\Phi_{\rm KN}$ and $\iota_{\rm EM}$ are essentially unconstrained by an optical light curves lacking the kilonova-characteristic color evolution. The evidence is hence diluted over this larger and uninformative prior volume even where the best-fit likelihood is comparable, so the rejection of the kilonova hypothesis is robust rather than an artifact of prior choice.

\begin{figure*}[ht!]
\centering
\includegraphics[width=\textwidth]{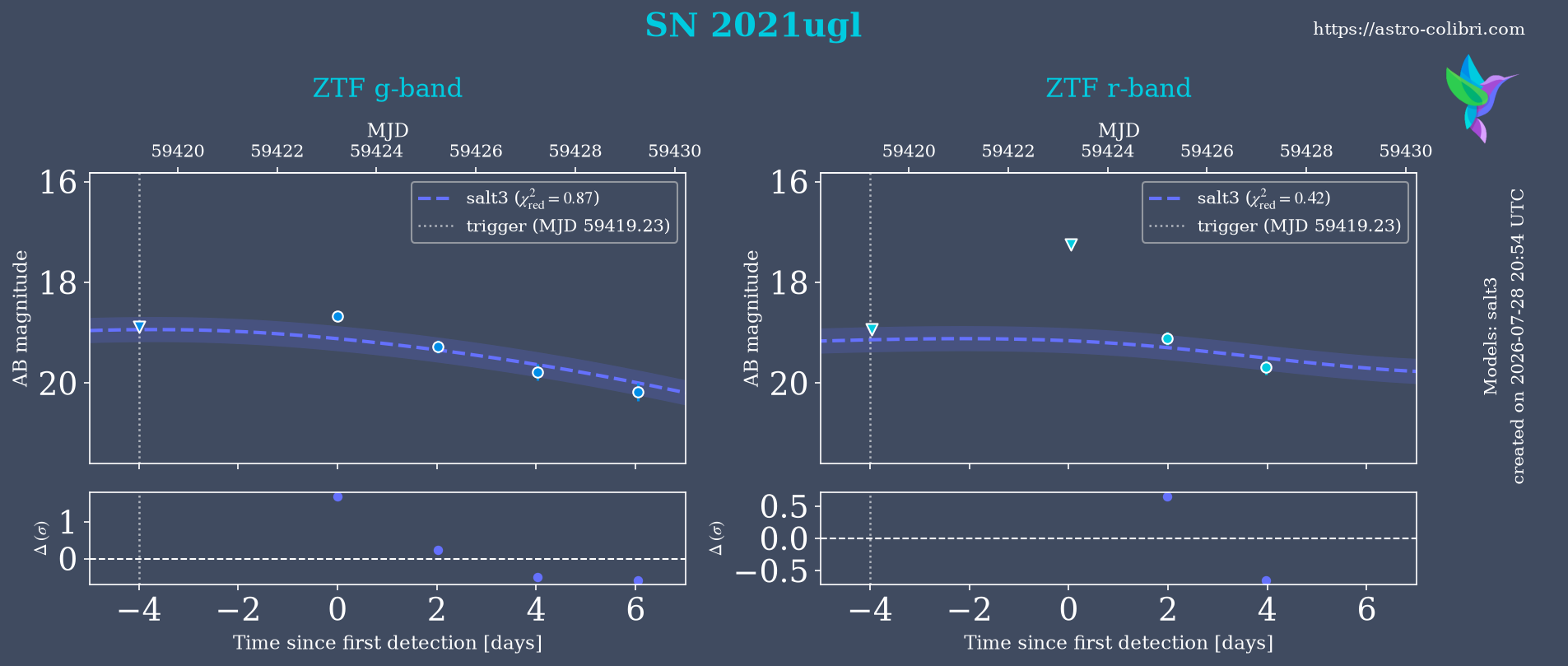}
\caption{Early-time \texttt{salt3} (Type\,Ia) fit to the first $\sim\!6$~days of SN\,2021ugl ZTF $g,r$ photometry, including the pre-trigger UL at $-4$~days (downward triangle) and the fitted trigger (dotted line, MJD~59419.23). Despite an apparently good match to the detections ($\chi^2_{\rm red}=0.87$ ($g$), $0.42$ ($r$)), \texttt{salt3} is decisively rejected by the Bayesian evidence ($\ln\mathcal{B}=11.69$; Table~\ref{tab:early}): a smooth monotonic curve cannot reproduce the shock-cooling decline sampled from $t=0$, and its three extra parameters dilute the evidence through the Occam factor. Goodness-of-fit and evidence rank this model very differently, the central methodological point of this work.}
\label{fig:lc_salt3}
\end{figure*}

\textbf{SN Type\,Ia}---The same evidence-versus-fit distinction sharpens the rejection of the Type\,Ia templates. Over the short early-time window, both \texttt{nugent-sn1a} and \texttt{salt3} achieve an acceptable per-band goodness-of-fit ($\chi^2_{\rm red}\lesssim1.8$; Figure~\ref{fig:lc_early}), and a visual inspection of their smooth, monotonic light curves might suggest a viable classification. They are nonetheless rejected by the evidence, at $\ln\mathcal{B}=4.45$ ($4.08$ with ULs) for \texttt{nugent-sn1a}, very strong on the Jeffreys scale, and decisively at $10.69$ ($11.69$ with ULs) for \texttt{salt3} (Table~\ref{tab:early}; Figure~\ref{fig:lc_salt3} shows the apparently good \texttt{salt3} fit). Three effects compound to produce this outcome. First, the per-filter $\chi^2_{\rm red}$ of Equation~\ref{eq:chi2} is evaluated on detections alone, so a smooth single-component template can pass within the error bars of every detection in a sparse light curve without incurring a large penalty, whereas the marginal evidence additionally integrates the survival-function term over the reported ULs (Equation~\ref{eq:likelihood}). Second, the decline sampled from $t=0$ is too rapid for a Type\,Ia template: over the early-time window \texttt{salt3} reproduces the post-maximum decay of a thermonuclear supernova, powered by $^{56}$Ni and evolving on a timescale of weeks (Figure~\ref{fig:lc_salt3}), and cannot match the day-timescale fade of shock-cooling emission that drives the classification. Third, \texttt{salt3} carries three additional free parameters ($x_0,x_1,c$; $N_\theta=6$ versus $4$ for the anchored templates), whose largely uninformative prior volume dilutes the evidence through the Occam factor, the same mechanism that penalizes the seven-parameter \texttt{Bu2019lm} above.

That a template with an acceptable $\chi^2$ is still strongly disfavored illustrates that robust photometric classification must rest on the marginal Bayesian evidence rather than on goodness-of-fit alone.

\begin{table*}[ht!]
\centering
\caption{Early-time (first $\sim\!6$~days, ZTF-only $g,r$) nested-sampling log Bayes factors and best-fit per-band reduced chi-squared, relative to the preferred \texttt{v19-1993j-corr} template, available on 2021~August~3, ten days before the spectroscopic confirmation, with and without the pre-trigger and same-night ULs. Models are ranked by the detections-only $\ln\mathcal{B}$ column; $\ln\mathcal{B} = \ln\mathcal{Z}_{\rm best} - \ln\mathcal{Z}$.}
\label{tab:early}

\begin{tabular}{lcccccc}
\hline\hline
& \multicolumn{2}{c}{$\ln\mathcal{B}$} & \multicolumn{2}{c}{$\chi^2_{\rm red}(g)$} & \multicolumn{2}{c}{$\chi^2_{\rm red}(r)$} \\
\cmidrule(lr){2-3} \cmidrule(lr){4-5} \cmidrule(lr){6-7}
Model & det.\ only & with ULs & det.\ only & with ULs & det.\ only & with ULs \\
\hline
\texttt{v19-1993j-corr} (SN\,IIb)          & $0.00$  & $0.00$  & $0.04$ & $0.04$ & $0.01$ & $0.01$ \\
\texttt{v19-2008d-corr} (SN\,Ib)           & $1.58$  & $3.55$  & $0.08$ & $0.42$ & $0.20$ & $0.84$ \\
\texttt{Piro2021} (Shock cooling)          & $2.90$  & $3.57$  & $0.38$ & $0.68$ & $0.03$ & $0.06$ \\
\texttt{nugent-sn2l} (SN\,IIL)             & $3.91$  & $4.13$  & $1.65$ & $1.64$ & $0.58$ & $0.58$ \\
\texttt{nugent-sn1a} (SN\,Ia)              & $4.45$  & $4.08$  & $1.74$ & $1.84$ & $0.71$ & $0.68$ \\
\texttt{nugent-sn2n} (SN\,IIn)             & $4.76$  & $4.28$  & $2.98$ & $2.81$ & $1.01$ & $0.88$ \\
\texttt{nugent-hyper} (SN\,Ib/c hypern.)   & $5.32$  & $5.69$  & $1.64$ & $1.74$ & $0.74$ & $0.71$ \\
\texttt{v19-1994i-corr} (SN\,Ic)           & $5.38$  & $6.70$  & $0.52$ & $1.47$ & $1.21$ & $0.03$ \\
\texttt{nugent-sn2p} (SN\,IIP)             & $5.68$  & $5.52$  & $2.35$ & $2.35$ & $0.79$ & $0.82$ \\
\texttt{nugent-sn1bc} (SN\,Ib/c)           & $6.60$  & $6.98$  & $1.76$ & $2.09$ & $2.10$ & $1.49$ \\
\texttt{Bu2019lm} (kilonova)               & $9.12$  & $9.83$  & $1.30$ & $1.12$ & $1.33$ & $1.67$ \\
\texttt{salt3} (SN\,Ia)                    & $10.69$ & $11.69$ & $0.19$ & $0.87$ & $0.45$ & $0.42$ \\
\hline
\end{tabular}

\vspace{2pt}
\begin{minipage}{\textwidth}
\footnotesize
Statistical uncertainties on individual $\ln\mathcal{Z}$ values are $\simeq 0.04$--$0.09$. \texttt{Bu2019lm} is decisively rejected in both configurations despite its seven free parameters; \texttt{salt3} achieves the best per-band $\chi^2_{\rm red}$ of the library (detections-only) yet is rejected most strongly of all models, illustrating the Occam-factor penalty discussed in Section~\ref{subsec:early}.
\end{minipage}
\end{table*}

The early-time hierarchy is not an artifact of the limited window: the same template, \texttt{v19-1993j-corr}, achieves the highest evidence in the full 47-day baseline (Section~\ref{subsec:fullbaseline}, Table~\ref{tab:Bayes_factor}), so the real-time classification available ten days before spectroscopy is confirmed by the complete light curve.

\subsubsection{Full-baseline classification}
\label{subsec:fullbaseline}
 
Table~\ref{tab:Bayes_factor} lists the nested-sampling log-evidences for the model library, fitted to the full 47-day, three-band photometry, with and without the ULs. The empirical Type\,IIb template \texttt{v19-1993j-corr} \citep{Vincenzi_2019}, achieves the highest evidence in both configurations. The runner-up in both cases is the Type\,Ib (\texttt{v19-2008d-corr}), the other stripped-envelope template of the library, disfavoured by $\ln\mathcal{B} = 12.95 \pm 0.1$ (with ULs) and $8.28 \pm 0.1$ (detections only), decisive evidence on the Jeffreys scale. Every other subclass (Type\,Ia, II-P, II-L, IIn, Ib/c, hypernova) is rejected by $\ln\mathcal{B} > 25$. The classification is therefore robust at two levels: the \emph{event} is unambiguously a stripped-envelope supernova, and within that subclass the SN\,1993J-like Type\,IIb template is decisively preferred.

\begin{table*}[ht!]
\centering
\caption{Full-baseline (47-day, three-band) nested-sampling log Bayes factors and best-fit per-band reduced chi-squared for ten of the eleven supernova models, with and without ULs (\texttt{Piro2021} excluded; see note). Models are ranked by the detections-only $\ln\mathcal{B}$ column; $\ln\mathcal{B} = \ln\mathcal{Z}_{\rm best} - \ln\mathcal{Z}$.}
\label{tab:Bayes_factor}
\footnotesize
\begin{tabular}{lcccccccc}
\hline\hline
& \multicolumn{2}{c}{$\ln\mathcal{B}$} & \multicolumn{2}{c}{$\chi^2_{\rm red}(g)$} & \multicolumn{2}{c}{$\chi^2_{\rm red}(r)$} & \multicolumn{2}{c}{$\chi^2_{\rm red}(o)$} \\
\cmidrule(lr){2-3} \cmidrule(lr){4-5} \cmidrule(lr){6-7} \cmidrule(lr){8-9}
Model & det.\ only & with ULs & det.\ only & with ULs & det.\ only & with ULs & det.\ only & with ULs \\
\hline
\texttt{v19-1993j-corr} (SN\,IIb)          & $0.00$  & $0.00$  & $0.22$ & $0.25$ & $0.21$ & $0.16$ & $0.18$ & $0.44$ \\
\texttt{v19-2008d-corr} (SN\,Ib)           & $8.28$  & $12.95$ & $1.13$ & $1.73$ & $0.61$ & $0.68$ & $0.28$ & $0.38$ \\
\texttt{nugent-sn2n} (SN\,IIn)             & $25.82$ & $50.92$ & $3.65$ & $2.89$ & $1.07$ & $1.27$ & $0.72$ & $1.73$ \\
\texttt{v19-1994i-corr} (SN\,Ic)           & $27.42$ & $65.35$ & $2.12$ & $3.29$ & $2.41$ & $2.13$ & $0.55$ & $1.51$ \\
\texttt{nugent-sn1a} (SN\,Ia)              & $27.88$ & $35.86$ & $4.97$ & $5.23$ & $0.90$ & $0.75$ & $0.33$ & $0.70$ \\
\texttt{nugent-sn2p} (SN\,IIP)             & $28.46$ & $48.93$ & $3.10$ & $3.68$ & $2.18$ & $2.10$ & $0.57$ & $1.26$ \\
\texttt{nugent-hyper} (SN\,Ib/c hypern.)   & $28.64$ & $38.91$ & $4.71$ & $5.58$ & $0.97$ & $0.75$ & $0.62$ & $1.02$ \\
\texttt{salt3} (SN\,Ia)                    & $29.15$ & $42.04$ & $3.25$ & $3.73$ & $0.93$ & $0.73$ & $0.20$ & $0.71$ \\
\texttt{nugent-sn1bc} (SN\,Ib/c)           & $37.54$ & $48.82$ & $5.84$ & $6.31$ & $1.65$ & $1.26$ & $0.66$ & $1.10$ \\
\texttt{nugent-sn2l} (SN\,IIL)             & $49.35$ & $92.19$ & $4.71$ & $10.63$& $3.82$ & $2.00$ & $0.86$ & $1.43$ \\
\hline
\end{tabular}

\vspace{2pt}
\begin{minipage}{\textwidth}
\footnotesize
Statistical uncertainties on individual $\ln\mathcal{Z}$ values are $\simeq 0.06$--$0.10$. The \texttt{Piro2021} shock-cooling model is excluded from this ranking: as a model of early-time envelope cooling, it cannot reproduce the radioactively powered secondary peak near day~20 or the subsequent decline, and over the full 47-day baseline it is disfavoured by $\ln\mathcal{B}\simeq3.2\times10^{3}$ (detections only: $3182$; with ULs: $3217$), a rejection so extreme it merely confirms the model is being applied outside its domain of validity. It is instead assessed on its natural, early-time domain in Section~\ref{subsec:early}, where it is a competitive fit.
\end{minipage}
\end{table*}

\begin{figure*}[ht!]
\centering
\includegraphics[width=\textwidth]{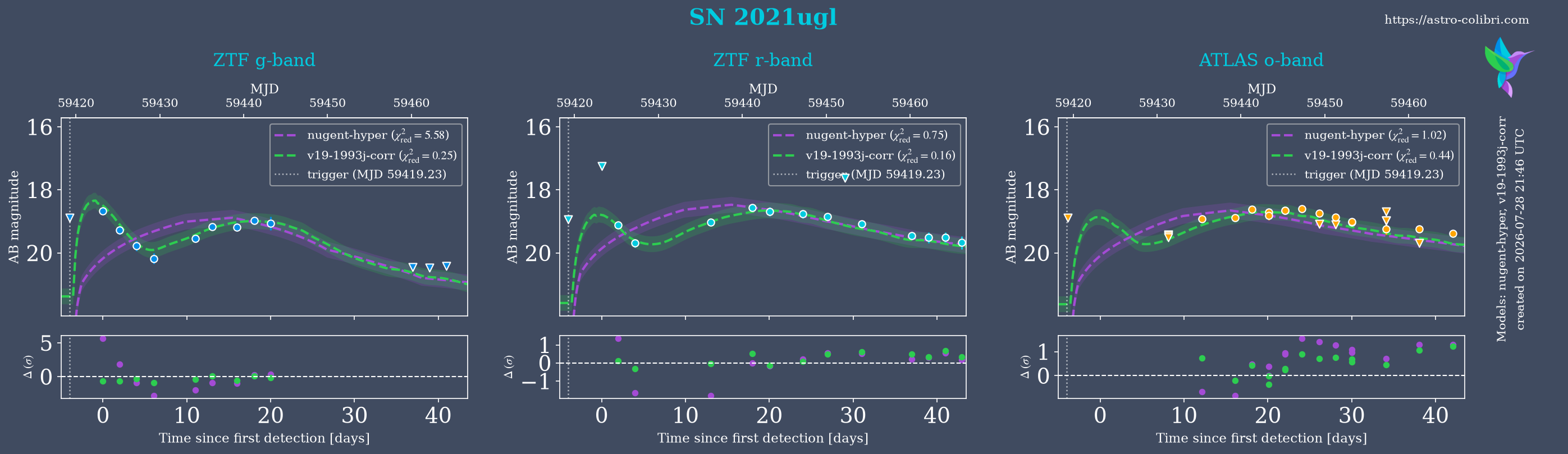}\\[10pt]
\includegraphics[width=\textwidth]{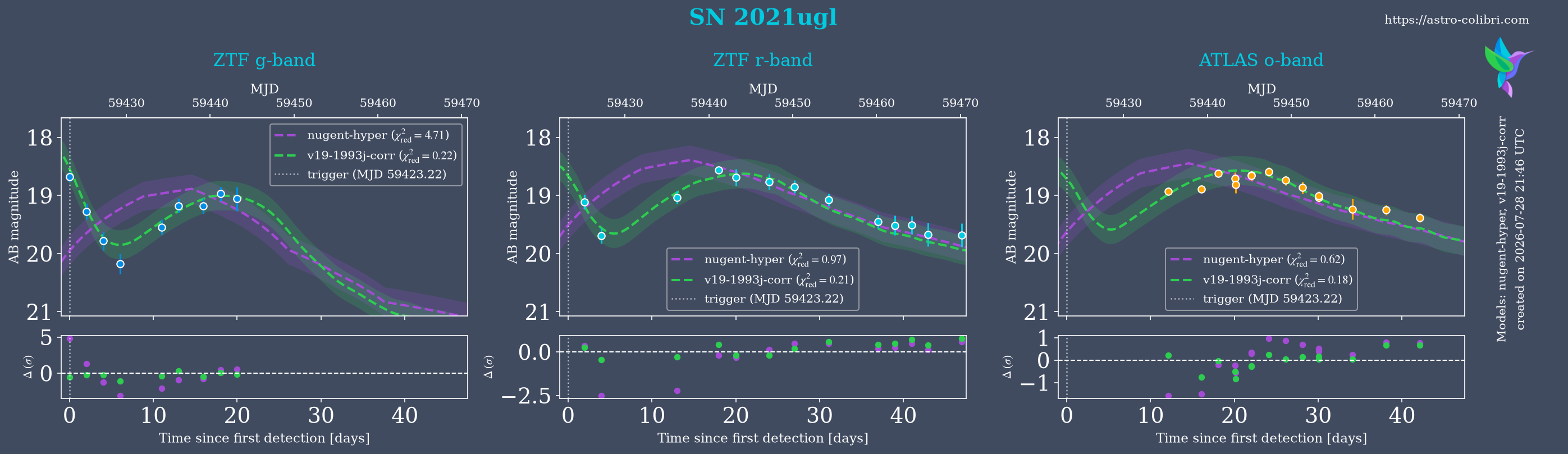}
    \caption{Best-fit light curves overlaying \texttt{v19-1993j-corr} (Type~IIb, preferred model) and \texttt{nugent-hyper} (best single-peaked alternative), fit to the full 47-day ZTF+ATLAS baseline. \textbf{Top}: full-baseline fit including the reported ULs (downward triangles) through the survival-function term of Equation~\ref{eq:likelihood}; \texttt{v19-1993j-corr} reproduces the early shock-cooling peak, the day-0 dip, and the radioactively powered maximum near day~20 in all three bands, while \texttt{nugent-hyper} misses the early peak entirely, missing the first ZTF~$g$-band detection by several~$\sigma$ and driving the $\ln\mathcal{B}=38.91$ evidence gap of Table~\ref{tab:Bayes_factor}. \textbf{Bottom}: detections-only fit, with the ULs removed, isolating their contribution to the model ranking; the preference for \texttt{v19-1993j-corr} persists ($\ln\mathcal{B}=28.64$), showing that the classification is driven by the shape of the detected light curve rather than by the ULs alone.}
    \label{fig:lc_full}
\end{figure*}

\begin{figure*}[ht!]
\centering
\includegraphics[width=\textwidth]{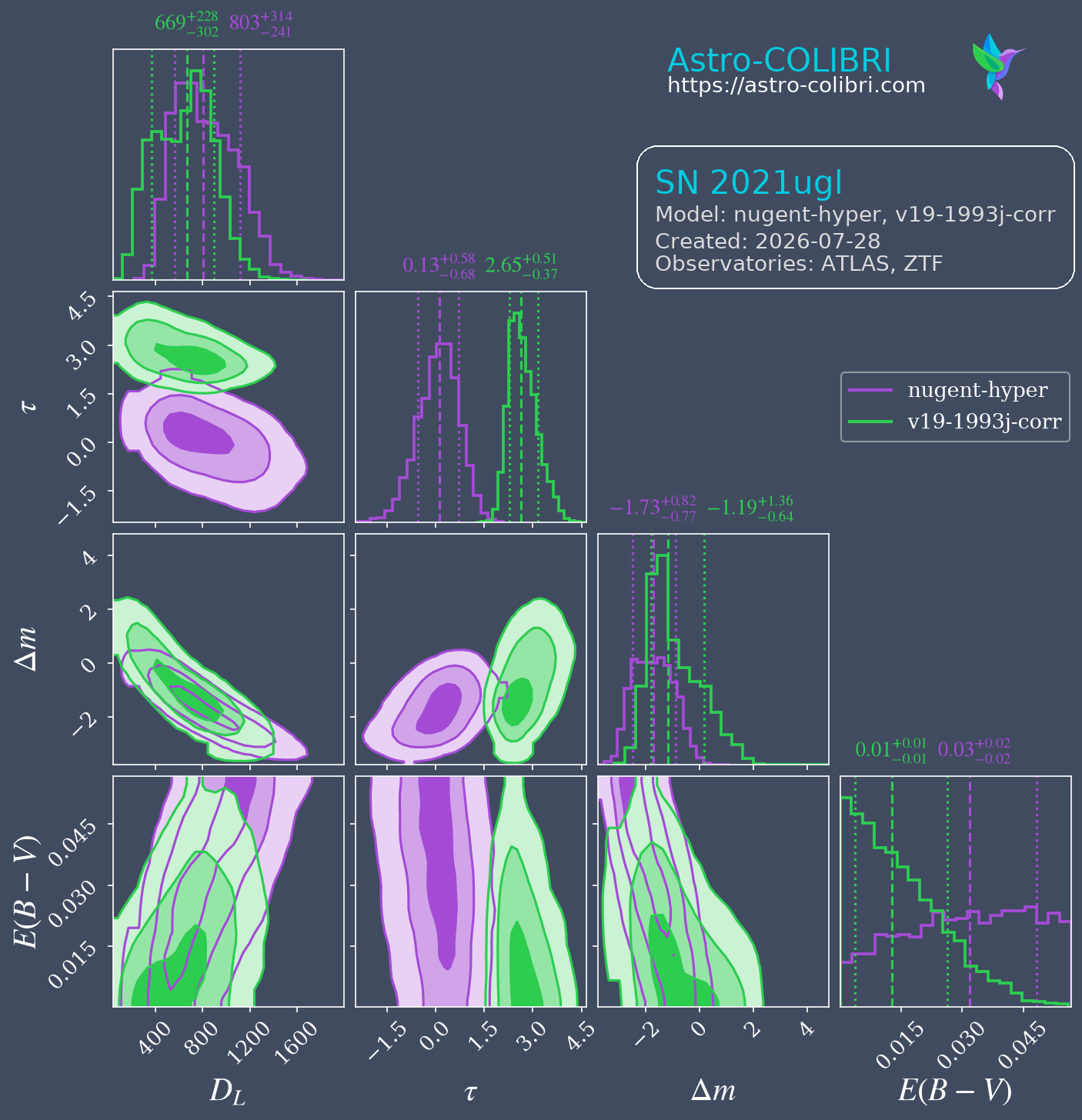}
\caption{Posterior distributions of the parameters shared by both models ($D_L$, $\tau$, $\Delta m$, $E(B-V)$) for the full-baseline (with ULs) fits of \texttt{v19-1993j-corr} (green) and \texttt{nugent-hyper} (magenta), with 1$\sigma$/2$\sigma$/3$\sigma$ credible contours. Dashed lines mark the median of each posterior; dotted lines mark the 16th/84th percentile bounds. Values above each 1D histogram give the median and asymmetric uncertainties, color-coded per model. The posteriors are well resolved within the prior volume, with the expected $D_L$-$\Delta m$ anti-correlation from the brightness degeneracy (Section~\ref{subsec:inference}).}
\label{fig:corner_full}
\end{figure*}

The best-fit \texttt{v19-1993j-corr} light curve (Figure~\ref{fig:lc_full}) reproduces the early shock-cooling peak, the dip coincident with the first detection, and the radioactively powered maximum near day~20 in all three bands and both configurations, with per-band reduced chi-squared values of $\chi^2_{\rm red}=0.25$ ($g$), $0.16$ ($r$), and $0.44$ ($o$) including the ULs ($0.22$, $0.21$, $0.18$ detections-only; Table~\ref{tab:Bayes_factor}). Values below unity reflect the conservative systematic floor $\sigma_{\rm eb} = 0.25$~mag; since the same floor is applied to every model, it does not affect the ranking. By contrast, the best of the single-peaked alternatives (\texttt{nugent-hyper}) fits the post-maximum decline in both configurations but misses the early shock-cooling peak entirely, failing the first $g$-band detection by several~$\sigma$, the morphological feature that drives the evidence gap. The corresponding posterior distributions for the shared parameters ($D_L$, $\tau$, $\Delta m$, $E(B-V)$) are shown in Figure~\ref{fig:corner_full}, which illustrates the expected $D_L$--$\Delta m$ anti-correlation from the brightness degeneracy.

\section{Discussion}
\label{sec:discussion}

\textbf{From goodness-of-fit to evidence}---The SN\,2021ugl analysis illustrates the central methodological premise of the service: robust photometric classification must rest on the marginal Bayesian evidence rather than on goodness-of-fit alone. Over the early-time window, the Type\,Ia templates \texttt{salt3} and \texttt{nugent-sn1a} reach formally acceptable per-band $\chi^2_{\rm red}$, yet are decisively rejected by the evidence (Section~\ref{subsec:early}). A smooth, single-component light curve can thread a sparse set of detections without a large $\chi^2$ penalty, but it neither reproduces the shock-cooling decline sampled from $t=0$ nor escapes the Occam factor that penalizes its additional free parameters. The same mechanism disfavors the seven-parameter kilonova template \texttt{Bu2019lm}. For this event, the service provides the discrimination a real-time pipeline requires: a transient flagged as a kilonova candidate is identified as a stripped-envelope supernova from photometry alone, ten days before the spectroscopic confirmation of \citet{Ridley_2021}.

\textbf{Limitations and future directions}---Three limitations frame the service. First, the computational cost of nested sampling, while modest for the four- to six-parameter supernova templates used here, grows for higher-dimensional models; keeping fits abreast of incoming photometry as the LSST alert stream reaches full scale will require further optimization. Second, the fits reported here marginalize over an unconstrained luminosity distance, so the brightness offset $\Delta m$ is not interpreted physically (Section~\ref{subsec:inference}); when a spectroscopic redshift is available, fixing $D_L$ restores a physical brightness measurement, and this mode is already supported. Third, the current library, though broad, is not exhaustive: rarer or peculiar subtypes fall outside it, and an event with no adequate template would be classified only relative to the models present.

In the longer term, the service will incorporate the \texttt{TrPi2018} GRB-afterglow model (Section~\ref{subsec:grb_model}), complementing the kilonova templates already available in the classification picker (Section~\ref{subsec:kn_models}). Early-time UV photometry from ULTRASAT \citep{shvartzvald2023ultrasat} and UVEX \citep{kulkarni2023science, Criswell_2025, Singer_2025} will further sharpen the discrimination between shock-cooling supernovae and kilonova candidates during gravitational-wave observing runs.

Noise regression techniques in the detectors \citep{Vajente_2020, Ormiston_2020, Saleem_2024, Kiendrebeogo_2025} improve strain sensitivity, alongside early-warning searches for binary neutron star inspirals \citep{Magee_2021} that allow telescopes to be pointed before coalescence. A known coalescence time would tighten the prior on the trigger-time offset $\tau$ (Table~\ref{tab:priors}), so that a candidate whose inferred explosion time falls outside the gravitational-wave window could be rejected from photometry alone.

\section{Conclusion}
\label{sec:conclusion}

We have presented \texttt{NMMA--Astro-COLIBRI}, an on-demand Bayesian light-curve classification service that couples the NMMA inference framework to the \emph{Astro-COLIBRI} real-time multi-messenger platform. A user can select a transient from the \emph{Astro-COLIBRI} archive and fit its multi-survey photometry against a library of eleven supernova models and five kilonova models, receiving best-fit light curves, corner plots, and Bayesian evidences within minutes. Running several models on the same event yields a ranking of competing subtypes through the log Bayes factor. The results are delivered directly to the web and mobile clients, reaching tens of thousands of professional and amateur users without requiring membership in a dedicated collaboration.

Demonstrated on SN\,2021ugl, a Type\,IIb supernova initially mistaken for a kilonova candidate, the service recovers the correct classification from the first $\sim\!6$~days of photometry alone, favoring the empirical Type\,IIb template \texttt{v19-1993j-corr} over the kilonova-mimicking shock-cooling model by $\ln\mathcal{B}=3.57$ and over the kilonova template \texttt{Bu2019lm} by $\ln\mathcal{B}=9.83$, a conclusion confirmed by the full 47-day baseline, where \texttt{v19-1993j-corr} remains preferred over its closest competitor, \texttt{v19-2008d-corr}, by $\ln\mathcal{B}=12.95$. This demonstrates that a complete supernova-subtype library, coupled to quantitative Bayesian model selection and delivered in real-time, is a prerequisite for credible kilonova discrimination in the multi-survey era.

\section*{\textbf{Software and Data Availability}}

\textbf{Astro-COLIBRI platform}---The \texttt{NMMA--Astro-COLIBRI} service is publicly accessible through the \emph{Astro-COLIBRI} platform at \url{https://astro-colibri.science}, with documentation at \url{https://nmma.live}. \\

\textbf{NMMA framework}---The NMMA framework itself is publicly available at \url{https://github.com/nuclear-multimessenger-astronomy/nmma}.\\

\textbf{Reproducibility repository}---The scripts, posterior samples, best-fit parameters, and figures needed to reproduce the SN\,2021ugl case study presented in Section~\ref{sec:results} are publicly available at \url{https://github.com/astro-transients/nmma-astrocolibri-sn2021ugl}.\\

\textbf{Archived data}---The data underlying this repository are archived on Zenodo at \url{https://doi.org/10.5281/zenodo.21771187}.\\

\textbf{Software packages}---This work made use of \texttt{NMMA} \citep{Pang_2023, Rose:2026}, \texttt{BILBY} \citep{Ashton:2018jfp,Romero-Shaw:2020owr}, \texttt{MultiNest} \citep{Feroz_2009} via \texttt{PyMultiNest} \citep{Buchner2014}, \texttt{SNCosmo} \citep{barbary2016sncosmo,barbary_2025_15019859}, and \texttt{corner} \citep{corner}.

\section{Acknowledgements}

R.W.K and F.S. acknowledge the ANR (French National Research Agency) for its support of the project ``Multi-messenger observations of the Transient Sky'' (MOTS), grant no.\ ANR-22-CE31-0012.

R.W.K. and A.SP. thanks the Action Thématique des Phénomènes Extrêmes et Multi-Messagers for its support.

We also thank the ACME project, funded by the European Union's Horizon Europe research and innovation programme under grant agreement no.\ 101131928.

We thank the Astro-COLIBRI team, past and present, for their work in developing and maintaining the platform.


\pagebreak
\newpage
\bibliographystyle{aasjournal}
\bibliography{references}

\end{document}

%% file: acronyms.tex
\providecommand{\acrolowercase}[1]{\lowercase{#1}}

\begin{acronym}
\acro{2D}[2D]{two\nobreakdashes-dimensional}
\acro{3D}[3D]{three\nobreakdashes-dimensional}
\acro{aLIGO}[aLIGO]{Advanced \acs{LIGO}}
\acro{AdVirgo}[AdVirgo]{Advanced Virgo}
\acro{ASD}[ASD]{amplitude spectral density}
\acro{BAYESTAR}[BAYESTAR]{BAYESian TriAngulation and Rapid localization}
\acro{BBH}[BBH]{binary black hole}
\acro{BH}[BH]{black hole}
\acro{BNS}[BNS]{binary neutron star}
\acro{CBC}[CBC]{compact binary coalescence}
\acro{CNN}[CNN]{convolutional neural network}
\acro{EM}[EM]{electromagnetic}
\acro{EOS}[EOS]{equation of state}
\acro{FAR}[FAR]{false alarm rate}
\acro{FOV}[FOV]{field of view}
\acroplural{FOV}[FOV\acrolowercase{s}]{fields of view}
\acro{GCN}[GCN]{Gamma-ray Coordinates Network}
\acro{GR}[GR]{general relativity}
\acro{GRANDMA}[GRANDMA]{Global Rapid Advanced Network Devoted to Multi-messenger Addicts}
\acro{GRB}[GRB]{gamma-ray burst}
\acro{GW}[GW]{gravitational-wave}
\acro{gwemopt}[gwemopt]{Gravitational-wave  Electromagnetic Optimization}
\acro{GWTC-3}[GWTC-3]{Gravitational Wave Transient Catalogue 3}
\acro{H0}[$H_0$]{Hubble-Lemaître constant} 
\acro{HEALPix}[HEALP\acrolowercase{ix}]{Hierarchical Equal Area isoLatitude Pixelization}
\acro{IGWN}[IGWN]{international gravitational-wave network}
\acro{ISCO}[ISCO]{innermost stable circular orbit}
\acro{KAGRA}[KAGRA]{KAmioka GRAvitational\nobreakdashes-wave observatory}
\acro{KDE}[KDE]{kernel density estimator}
\acro{KN}[KN]{kilonova}
\acroplural{KN}[KNe]{kilonovae}
\acro{LHO}[LHO]{\ac{LIGO} Hanford Observatory}
\acro{LIGO}[LIGO]{Laser Interferometer \acs{GW} Observatory}
\acro{LISA}[LISA]{The Laser Interferometer Space Antenna} 
\acro{LLO}[LLO]{\ac{LIGO} Livingston Observatory}
\acro{LRR}[LRR]{Living Reviews in Relativity}
\acro{LSC}[LSC]{\acs{LIGO} Scientific Collaboration}
\acro{LSST}[LSST]{Large Synoptic Survey Telescope}
\acro{Msun}[\ensuremath{M_{\odot}}]{sun mass}
\acro{MLE}[MLE]{\ac{ML} estimator}
\acro{ML}[ML]{maximum likelihood}
\acro{NIR}[NIR]{near infrared}
\acro{NMMA}[NMMA]{Nuclear-physics and Multi-Messenger Astrophysics}
\acro{NSBH}[NSBH]{neutron star\nobreakdashes--black hole}
\acro{NSBH}[NSBH]{\acl{NS}\nobreakdashes--\acl{BH}}
\acro{NS}[NS]{neutron star}
\acro{O1}[O1]{\ac{LIGO}'s first observing run}
\acro{O2}[O2]{\ac{LIGO}/Virgo's second observing run}
\acro{O3}[O3]{\ac{LIGO}/Virgo's third observing run}
\acro{O4}[O4]{\acs{LIGO}/Virgo/\acs{KAGRA}'s fourth observing run}
\acro{O5}[O5]{\acs{LIGO}/Virgo/\acs{KAGRA}'s fifth observing run}
\acro{O6}[O6]{\acs{LIGO}/Virgo/\acs{KAGRA}/LIGO\nobreakdashes-India's fifth observing run}
\acro{PSD}[PSD]{power spectral density}
\acro{PDB}[PDB]{Power Law + Dip + Break}
\acro{PDB/GWTC-3}[PDB/GWTC-3]{\acl{PDB}/\acs{GWTC-3}}
\acro{SED}[SED]{spectral energy distribution}
\acro{SN}[SN]{supernova}
\acroplural{SN}[SN\acrolowercase{e}]{supernovae}
\acro{SNII}[\acs{SN}\,II]{Type~II \ac{SN}}
\acroplural{SNII}[\acsp{SN}\,II]{Type~II \acp{SN}}
\acro{SNIa}[\acs{SN}\,I\acrolowercase{a}]{Type~Ia \ac{SN}}
\acroplural{SNIa}[\acsp{SN}\,I\acrolowercase{a}]{Type~Ia \acp{SN}}
\acro{SNIIb}[\acs{SN}\,II\acrolowercase{b}]{Type~IIb \ac{SN}}
\acroplural{SNIIb}[\acsp{SN}\,II\acrolowercase{b}]{Type~IIb \acp{SN}}
\acro{SNIb}[\acs{SN}\,I\acrolowercase{b}]{Type~Ib \ac{SN}}
\acroplural{SNIb}[\acsp{SN}\,II\acrolowercase{b}]{Type~Ib \acp{SN}}
\acro{SNIc}[\acs{SN}\,I\acrolowercase{c}]{Type~Ic \ac{SN}}
\acroplural{SNIc}[\acsp{SN}\,II\acrolowercase{c}]{Type~Ic \acp{SN}}
\acro{SNIbc}[\acs{SN}\,I\acrolowercase{b}/I\acrolowercase{c}]{Type~Ib/c \ac{SN}}
\acroplural{SNIbc}[\acsp{SN}\,I\acrolowercase{b}/I\acrolowercase{c}]{Type~Ib/c \acp{SN}}
\acro{SNR}[S/N]{signal\nobreakdashes-to\nobreakdashes-noise ratio}
\acro{SVD}[SVD]{singular value decomposition}
\acro{TOO}[TOO]{target\nobreakdashes-of\nobreakdashes-opportunity}
\acroplural{TOO}[TOO\acrolowercase{s}]{targets of opportunity}
\acro{ULTRASAT}[ULTRSAT]{Ultraviolet Transient Astronomy Satellite}
\acro{UV}[UV]{ultraviolet}
\acro{UVEX}[UVEX]{Ultraviolet Explorer}
\acro{Roman}[Roman]{Nancy Grace Roman Space Telescope}
\acro{Rubin}[Rubin Observatory]{Vera C.\ Rubin Observatory}
\acro{WD}[WD]{white dwarf}
\acro{ZTF}[ZTF]{Zwicky Transient Facility}
\end{acronym}